\documentstyle[psfig,macros,xcolor,soul]{mn}
\begin{document}
%

\title[Segregation of Substructure]
      {On the Segregation of Dark Matter Substructure}

\author[van den Bosch et al.]
       {\parbox[t]{\textwidth}{
        Frank C. van den Bosch$^{1}$\thanks{E-mail: frank.vandenbosch@yale.edu}, 
        Fangzhou Jiang$^{1}$,
        Duncan Campbell$^{1}$,
        Peter Behroozi$^{2}$}
           \vspace*{3pt} \\
        $^1$Department of Astronomy, Yale University, PO. Box 208101,
            New Haven, CT 06520-8101\\
        $^2$Space Telescope Science Institute, Baltimore, MD 21218, USA}


\date{}

\pagerange{\pageref{firstpage}--\pageref{lastpage}}
\pubyear{2013}

\maketitle

\label{firstpage}


\begin{abstract}
  We present the first comprehensive analysis of the segregation of
  dark matter subhaloes in their host haloes. Using three different
  numerical simulations, and two different segregation strength
  indicators, we examine the segregation of twelve different subhalo
  properties with respect to both orbital energy and halo-centric
  radius (in real space as well as in projection).  Subhaloes are
  strongly segregated by accretion redshift, which is an outcome of
  the inside-out assembly of their host haloes. Since subhaloes that
  were accreted earlier have experienced more tidal stripping,
  subhaloes that have lost a larger fraction of their mass at infall
  are on more bound orbits. Subhaloes are also strongly segregated in
  their masses and maximum circular velocities at accretion. We
  demonstrate that part of this segregation is already imprinted in
  the infall conditions. For massive subhaloes it is subsequently
  boosted by dynamical friction, but only during their first radial
  orbit. The impact of these two effects is counterbalanced, though,
  by the fact that subhaloes with larger accretion masses are accreted
  later. Because of tidal stripping, subhaloes reveal little to no
  segregation by present-day mass or maximum circular velocity, while
  the corresponding torques cause subhaloes on more bound orbits to
  have smaller spin.  There is a weak tendency for subhaloes that
  formed earlier to be segregated towards the center of their host
  halo, which is an indirect consequence of the fact that (sub)halo
  formation time is correlated with other, strongly segregated
  properties.  We discuss the implications of our results for the
  segregation of satellite galaxies in galaxy groups and clusters.
\end{abstract} 




\section{Introduction} 
\label{sec:intro}

It is well established that galaxy groups and clusters reveal
segregation of their member galaxies. In particular, early-type
galaxies (red, passive, elliptical) follow a more centrally
concentrated radial distribution (and have a narrower line-of-sight
velocity distribution) than their late-type (blue, active, spiral)
counterparts (e.g. Postman \& Geller 1984; Whitmore, Gilmore \& Jones
1993; Biviano \etal 1996, 2002; Carlberg \etal 1997; Balogh \etal
2000, 2004; Dom\'inguez \etal 2002; Girardi \etal 2003; G\'omez \etal
2003; Goto \etal 2003, 2004; Lares, Lambas \& S\'anchez 2004; Weinmann
\etal 2006; Blanton \& Berlind 2007; van den Bosch \etal 2008; Wetzel,
Tinker \& Conroy 2012).  In addition, numerous studies have detected
luminosity and/or stellar mass segregation in the sense that more
luminous and massive galaxies are located at smaller group- or
cluster-centric distances (e.g., Rood \& Turnrose 1968; Quintana 1979;
den Hartog \& Katgert 1996; Adami, Biviano \& Mazure 1998; Biviano
\etal 2002; Lares, Lambas \& S\'anchez 2004; McIntosh \etal 2005; van
den Bosch \etal 2008; Presotto \etal 2012; Balogh \etal 2014; Roberts
\etal 2015). However, it is important to point out that the presence
of luminosity and/or stellar mass segregation in groups and clusters
is still contentious (see e.g., Pracy \etal 2005; Hudson \etal 2010;
von der Linden \etal 2010; Wetzel \etal 2012; Ziparo \etal 2013;
Vulcani \etal 2013).

What is the origin of this segregation? It is useful to distinguish
the following two options. Segregation can arise as a manifestation of
the environmental impact of galaxy evolution. In this case, the
evolution of a satellite galaxy is directly (and causally) influenced
by its actual location (or orbit) in the dark matter halo of its
host. An example is ram-pressure stripping, which can cause a
satellite galaxy to quench its star formation. For ram-pressure to be
effective, the satellite needs to be moving relatively fast, through a
relatively dense gaseous atmosphere, which selects out orbits with a
small pericentric distance. Alternatively, segregation can also have
its origin in the dark sector, combined with the fact that galaxy
properties are tightly correlated with (sub)halo properties such as
mass. In this picture, there is segregation of dark matter subhaloes
according to one or more of their properties, while the galaxy
properties are determined by the pre-infall properties of their dark
matter haloes. An example is dynamical friction, which causes the
segregation of more massive subhaloes towards the center of the host
halo. If more massive subhaloes host more luminous galaxies, as
expected, luminosity segregation would be a natural outcome.  The main
difference between these two scenarios is whether the relevant galaxy
property (i.e., the property that reveals segregation) is determined
pre- or post-infall. Although it is useful to make this distinction,
we emphasize that these two `scenarios' are not mutually exclusive.

The origin of the segregation by galaxy type (color, star formation
rate, morphology) is most commonly linked to an environmental impact
of galaxy evolution, even though it is still heavily debated which
physical processes dominate the quenching and/or morphological
transformation of satellite galaxies. It is important, though, to
realize that this is not the only explanation. In fact, one can also
explain segregation by galaxy type as a manifestation of segregation
in the dark sector.  This was nicely demonstrated in Watson \etal
(2015), who showed that simply linking the star formation rate
(hereafter SFR) of a (satellite) galaxy to the formation time of its
dark matter halo, a procedure dubbed `age-matching' (Hearin \& Watson
2013; Hearin \etal 2014), can accurately reproduce the difference in
the radial distributions of quenched and active satellite galaxies in
host haloes that span several orders of magnitude in mass. This
suggests that dark matter subhaloes are somehow segregated by
formation time, and that this segregation is sufficient to explain the
segregation by SFR of their associated satellite galaxies.

The origin of luminosity segregation is most commonly linked to
dynamical friction, which implies that it has its origin in the dark
sector. However, this picture is complicated by the fact that dark
matter subhaloes experience tidal mass loss. Although more massive
subhaloes experience stronger dynamical friction, which causes them to
segregate out to smaller halo-centric distances, this also enhances
their mass loss rate, which in turn makes them less susceptible to
dynamical friction. Because of this tug of war between dynamical
friction and tidal mass stripping, it is difficult to predict a priori
whether and how subhaloes are segregated by mass. 

It is clear from the discussion above that understanding the origin of
segregation requires a detailed study as to how dark matter subhaloes
are segregated. In particular, in this day and age of linking galaxy
properties to dark matter properties, understanding segregation in the
dark sector is of paramount importance for understanding small-scale
clustering. Although there is no shortage of papers that have used
numerical simulations to study the statistics and properties of dark
matter subhaloes, there has been relatively little focus on how
subhaloes are segregated. Regarding mass segregation, simulations have
produced mixed results: whereas several authors find more massive
subhaloes to be {\it less} centrally concentrated in terms of their
radial distribution in the host halo (De Lucia \etal 2004; Reed \etal
2005; Angulo \etal 2009; Contini, De Lucia \& Borgani 2012), there are
also reports of no (significant) segregation by mass (Diemand, Moore
\& Stadel 2004; Springel \etal 2008; Ludlow \etal 2009), or of a weak
segregation in the opposite sense (Gao \etal 2004). Several studies
have pointed out that subhaloes selected based on their mass (or
maximum circular velocity) {\it at accretion} are more centrally
concentrated than those selected based on their present-day mass or
velocity (Gao \etal 2004; Nagai \& Kravtsov 2005; Kuhlen, Diemand \&
Madau 2007; Ludlow \etal 2009; Contini \etal 2012), although none of
these studies has investigated this in any detail. In terms of
properties other than mass, Gao \etal (2004), Faltenbacher \& Diemand
(2006) and Contini \etal (2012) have shown that subhaloes that were
accreted later are located at larger halo-centric radii, Wu \etal
(2013) find that subhaloes with larger maximum circular velocities are
segregated towards the center of their host halo, while Reed \etal
(2005) and Onions \etal (2013) find subhaloes at smaller halo-centric
radii to have lower spin.

This paper presents the first comprehensive study of the segregation
of dark matter subhaloes in numerical simulations. In particular, we
study the segregation of 12 different subhalo properties with respect
to both halo-centric radius (in real space as well as in projection)
and orbital energy. We compare the results from three different
simulations run with two different $N$-body codes, quantify
segregation strength using two different indicators, and carefully
analyze the origin of the various forms of segregation we
identify. There are three `mechanisms' than can give rise to
segregation of subhaloes. The most well-known is two-body relaxation,
which drives the system towards equipartition, and thus segregation by
mass. In the case of subhaloes, this two-body relaxation manifests
itself in the form of dynamical friction. Segregation may also arise
from the fact that dark matter haloes assemble from the inside out. As
a consequence, subhaloes that were accreted earlier, when the main
progenitor of the present day host halo was smaller, typically end up
at smaller halo-centric distances.  Any subhalo property that is
correlated with accretion time is therefore likely to show some amount
of segregation. And finally, it may be the case that there are
correlations between some subhalo properties and their orbital
properties at accretion. In particular, if subhalo property $P$ is
correlated with the orbital energy at infall, segregation with respect
to property $P$ will have been imprinted {\it at accretion}.  In
addition to these three `primary' mechanisms, segregation can also
have an indirect origin, and simply arise from a correlation with
another property that is physically segregated. In this paper we will
investigate which subhalo properties are segregated and with what
strength, and we will also examine which of the various mechanisms
discussed above is responsible for this segregation. In a forthcoming
paper (Lu et al., in preparation), we use this knowledge combined with
the observed segregation of galaxy properties such as luminosity,
stellar mass, color and star formation rate, to put constraints on the
galaxy-dark matter connection and on the various physical processes
related to galaxy formation and evolution.
\begin{table*}\label{tab:simdata}
\caption{Numerical Simulations used in this Paper}
\begin{center}
\begin{tabular}{lccccccrrcl}
\hline\hline
 Simulation & $\Omega_{\rmm,0}$ & $\Omega_{\Lambda,0}$ & $\Omega_{\rmb,0}$ & $\sigma_8$ & $n_\rms$ & $h$ & $L_{\rm box}$ & $N_\rmp$ & $m_\rmp$ & Reference \\
  (1) & (2) & (3) & (4) & (5) & (6) & (7) & (8) & (9) & (10) & (11) \\
\hline
Bolshoi      & 0.270 & 0.730 & 0.047 & 0.82 & 0.95 & 0.70 &   $250$ & $2048^3$ & $1.35 \times 10^8$ & Klypin \etal (2011) \\ 
Chin250      & 0.286 & 0.714 & 0.047 & 0.82 & 0.96 & 0.70 &   $250$ & $2048^3$ & $1.44 \times 10^8$ & Becker \etal (in prep.) \\ 
Chin400      & 0.286 & 0.714 & 0.047 & 0.82 & 0.96 & 0.70 &   $400$ & $2048^3$ & $5.91 \times 10^8$ & Becker \etal (in prep.) \\ 
\hline\hline
\end{tabular}
\end{center}
\medskip
\begin{minipage}{\hdsize}
  Parameters of the various numerical simulations used in this paper.
  Columns (2) - (7) list the present-day cosmological density
  parameters for the matter, $\Omega_{\rmm,0}$, the cosmological
  constant, $\Omega_{\Lambda,0}$, and the baryonic matter,
  $\Omega_{\rmb,0}$, the normalization, $\sigma_8$, and spectral
  index, $n_\rms$, of the matter power spectrum, and the Hubble
  parameter, $h = H_0/(100\kmsmpc)$. Columns (8) - (10) list the box
  size of the simulation, $L_{\rm box}$, (in $h^{-1} \Mpc$), the
  number of particles used, $N_\rmp$, and the particle mass, $m_\rmp$
  (in $h^{-1} \Msun$), respectively.  More details regarding each
  simulation can be found in the references listed in Column~(11).
\end{minipage}
\end{table*}

This paper is organized as follows. \S\ref{sec:Method} describes the
numerical simulations used, the selection of our sample of subhaloes,
the various segregation properties and indicators, as well as the two
statistics we use to gauge segregation strength. Results on the
segregation of our 12 subhalo properties are presented in
\S\ref{sec:res}, while their origin is discussed in \S\ref{sec:orig}.
We summarize and discuss our results in \S\ref{sec:disc}.


\section{Methodology}
\label{sec:Method}

This section describes the numerical simulations, the selection of
subhaloes, and the method used to assess their segregation. However,
we start with a brief introduction of halo basics, outlining a number
of definitions and notations.

\subsection{Halo Basics and Notation}
\label{sec:def}

Throughout this paper we distinguish between dark matter host haloes
and dark matter subhaloes. The latter have their center located
inside the virial radius of a host halo. Host haloes at redshift $z$ are
defined as spherical systems with a virial radius $r_{\rm vir}$ inside
of which the average density is equal to $\Delta_{\rm vir}(z) \,
\rho_{\rm crit}(z)$. Here $\rho_{\rm crit}(z) = 3 H^2(z)/8 \pi G$ is
the critical density for closure, and
\begin{equation}\label{deltavir}
\Delta_{\rm vir}(z) = 18\pi^2 + 82 x - 39 x^2\,
\end{equation}
with $x =\Omega_\rmm(z) - 1$ (Bryan \& Norman 1998).  The (virial)
mass of a host halo is defined as the mass within the virial radius
$r_{\rm vir}$ and indicated by $M$. Subhaloes are defined as haloes
whose center falls inside the virial radius of a (more massive) host
halo, and their mass, indicated by $m$, is defined as the sum of the
masses of the particles that are deemed bound to the subhalo by the
subhalo finder used (see below).

Throughout this paper, we approximate that host haloes follow an NFW
density profile (Navarro, Frenk \& White 1997), for which the
gravitational potential is given by
\begin{equation}\label{PhiNFW}
\Phi(r) = -V^2_{\rm vir} \, {\ln(1+cx) \over f(c) \, x} =
- \left({V_{\rm max} \over 0.465}\right)^2 \, {\ln(1+cx) \over cx}\,.
\end{equation}
Here $V_{\rm vir} = \sqrt{G M/r_{\rm vir}}$ is the circular velocity
at the virial radius, $x=r/r_{\rm vir}$,
\begin{equation}\label{fx}
f(x) = \ln(1+x) - {x \over 1+x}\,,
\end{equation}
and $c$ is the halo's concentration parameter.  The central potential
of an NFW halo is given by
\begin{equation}
\Phi_0 \equiv \Phi(0) = -\left({V_{\rm max} \over 0.465}\right)^2\,.
\end{equation}
where $V_{\rm max}$ is the maximum circular velocity, which for an
NFW profile is given by
\begin{equation} \label{VmaxHost}
V_{\rm max} = 0.465 \, V_{\rm vir} \, \sqrt{c \over f(c)}\,,
\end{equation}
We will use $\Phi_0$ throughout to normalize the orbital energy
of subhaloes. 

Finally, the (dimensionless) spin parameter of a dark matter (sub)halo
of mass $M$ is defined as
\begin{equation} \label{spin}
\lambda \equiv {J |E|^{1/2} \over G M^{5/2}}\,,
\end{equation}
(Peebles 1969). Here $J$ is the total angular momentum of the halo,
$G$ is the gravitational constant, and $E = T + U$ is the total
energy, with $T$ and $U$ the total kinetic and potential energies,
respectively.

\subsection{Numerical Simulation}
\label{sec:SIM}

In this paper we use three cosmological $N$-body simulations to study
the segregation of dark matter subhaloes: the `Bolshoi' simulation
(Klypin, Trujillo-Gomez \& Primack 2011), and two simulations
(`Chin250' and `Chin400') of the Chinchilla suite (Becker et al., in
prep.). The cosmological parameters of the Bolshoi and Chinchilla
simulations (see Table~1) are slightly different, but both are
consistent with the latest constraints from the Planck satellite
(Planck Collaboration 2015). Bolshoi has been run with the Adaptive
Refinement Tree ({\tt ART}) code (Kravtsov, Klypin \& Khokhlov 1997),
while the Chinchilla simulations have been run using {\tt Gadget-2}
(Springel 2005). All simulations follow the evolution of $2048^3$ dark
matter particles in a periodic simulation box. The box sizes and
particle masses of the simulations are listed in Table~1.

For all three simulations we use the $z=0$ halo catalogs obtained with
the phase-space halo finder \Rockstar (Behroozi \etal 2013a,b), which
uses adaptive, hierarchical refinement of friends-of-friends groups in
six phase-space dimensions and one time dimension. As demonstrated in
Knebe \etal (2011, 2013), this results in a very robust tracking of
(sub)haloes (see also van den Bosch \& Jiang 2014). In line with the
halo definition given above, \Rockstar host haloes are defined as
spheres with an average density equal to $\Delta_{\rm vir} \rho_{\rm
  crit}$. Subhaloes are defined as haloes whose center lies within the
virial radius, $r_{\rm vir}$, of a bigger halo. 
\begin{figure*}
\centerline{\psfig{figure=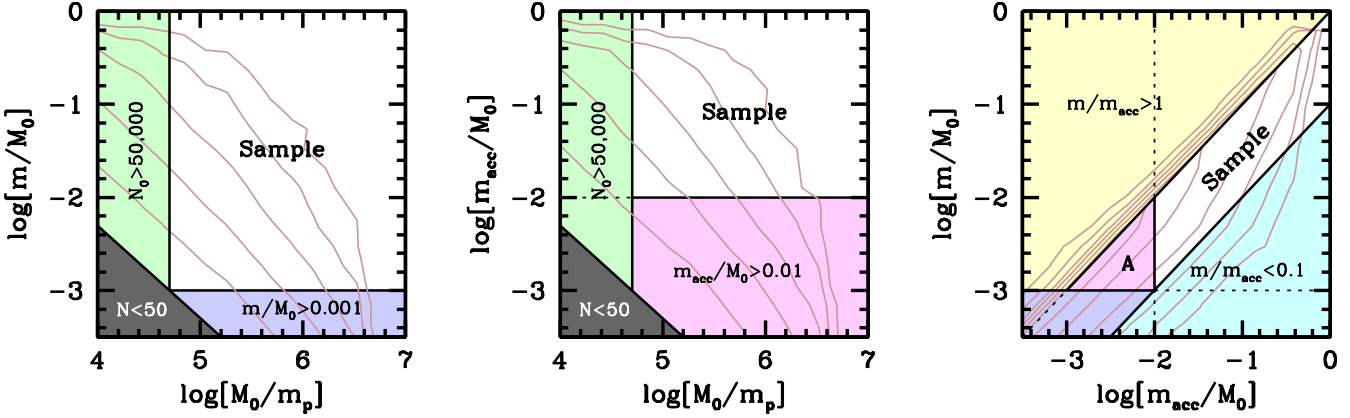,width=\hdsize}}
\caption{Illustration of the various criteria used for the selection
  of our sample of subhaloes. From left to right the panels plot
  $\log[m/M_0]$ {\it vs.}  $\log[M_0/m_\rmp]$, $\log[m_{\rm acc}/M_0]$
  {\it vs.}  $\log[M_0/m_\rmp]$, and $\log[m/M_0]$ {\it vs.}
  $\log[m_{\rm acc}/M_0]$. The white region, labeled `Sample'
  indicates the region in $(M_0,m,m_{\rm acc})$-parameter space
  occupied by subhaloes in our fiducial sample. Color shaded regions
  reflect the various cuts discussed in the text, while contours
  indicate the number density of subhaloes in the Bolshoi simulation.
  The triangular, pink-shaded region labeled `A' indicates the region
  in $(M_0,m,m_{\rm acc})$-parameter space that is added to our sample
  if we remove the constraint that $m_{\rm acc}/M_0 \geq 0.01$. The
  impact of adding these subhaloes to our sample is discussed
  throughout the text and in Appendix~\ref{App:sample}.  Note that we
  deliberately refrained from using primary colors in order to avoid a
  too close resemblance with the copyrighted art by Piet Mondrian.}
\label{fig:sample}
\end{figure*}

\subsection{Sample Selection}
\label{sec:sample}

When investigating the segregation of dark matter subhaloes in
numerical simulations, it is important to select your sample
carefully. Fig.~\ref{fig:sample} illustrates the various selection
criteria we have applied to the host haloes and subhaloes in our three
simulation boxes. First off, throughout we limit our study to
subhaloes with at least 50 particles, which is roughly the limit below
which resolution effects start to affect completeness in the subhalo
mass function (van den Bosch \& Jiang 2014).  The dark-grey triangular
region in the left-hand panel of Fig.~\ref{fig:sample} show the
constraints this puts on the ratio of the present-day subhalo mass,
$m$, to that of its host halo, $M_0$, as a function of $M_0/m_\rmp$,
where $m_\rmp$ is the simulation particle mass.  As a compromise
between sample size and dynamical range, we first construct a complete
sample of all subhaloes with $\log[\msub] \geq -3$, which excludes the
purple-shaded region in Fig.~\ref{fig:sample}.  Having the sample be
complete implies that we restrict ourselves to host haloes with at
least $50,000$ particles, which excludes the green-shaded regions in
Fig.~\ref{fig:sample}. This constraint on host halo mass translates to
$M_0 > 6.75 \times 10^{12} \Msunh$ for the Bolshoi simulation, while
the equivalent lower limits for the Chin250 and Chin400 simulations
are $7.20 \times 10^{12} \Msunh$ and $2.96 \times 10^{13} \Msunh$,
respectively.

In addition to these constraints on $m$ and $M_0$, we also select our
subhaloes based on their mass at accretion, $m_{\rm acc}$.  Since one
typically expects a subhalo to loose mass after accretion, due to
tidal stripping by the host halo, we demand that $\mrat \leq 1$,
excluding the yellow-shaded region in the right-hand panel. This only
removes $\sim 5\%$ of the subhaloes, most of which have a present day
subhalo mass just slightly larger than unity. We also exclude
subhaloes that have lost more than 90 percent of their mass, which
excludes the cyan-shaded region in the right-hand panel of
Fig.~\ref{fig:sample}.  We emphasize that removing these two
constraints on $\mrat$ (while keeping all other constraints intact)
only increases the sample size by $\sim 13\%$, and has no discernible
impact on any of the results presented below. We mainly use these
criteria to have a sharper defined sample.

Finally, we also impose that $\log[\macc] \geq -2$, which excludes the
pink-shaded regions in Fig.~\ref{fig:sample}. This is done to have a
sample of subhaloes that is more reminiscent of what observers may use
to study segregation in galaxy groups and clusters. After all, the
luminosities and stellar masses of satellite galaxies are believed to
be tightly correlated with the corresponding subhalo masses {\it at
  accretion}, which is the premise of the succesful and popular
subhalo abundance matching technique (e.g., Vale \& Ostriker 2006;
Conroy, Wechsler \& Kravtsov 2006; Behroozi, Conroy \& Wechsler
2010). This cut in $\macc$ severely reduces the sample size, by about
a factor four. In particular, in the right-hand panel of
Fig.~\ref{fig:sample} there are roughly three times as many subhaloes
in the pink, triangular region (marked `A') than in the white region
demarcating our fiducial sample. As we discuss in detail in
Appendix~\ref{App:sample}, adding these galaxies to our sample (i.e.,
removing the constraint that $\log[\macc] \geq -2$) has a strong
impact on the segregation of subhaloes by present-day mass (see also
\S\ref{sec:massresults} and \S\ref{sec:mass}).

In addition to the constraints on $M_0$, $m$, and $m_{\rm acc}$, we
also remove all subhaloes ($\sim 2.7\%$) for which the virial ratio
$T/|U| > 1$ (see \S\ref{sec:properties} below for a definition).
These latter are subhaloes that are very far from virial equilibrium,
and for which masses and circular velocities are extremely uncertain.
We emphasize, though, that this restriction again has no significant
impact on any of our results. Finally, we remove all subhaloes ($\sim
1.1\%$) whose orbital energy $E > 0$, and which are thus not bound to
their host. For each of the three simulation boxes, this yields
samples of $\sim 22,000$ subhaloes, for a grand total of $66,401$
subhaloes in all three simulation boxes combined.

We acknowledge that these selection criteria are somewhat arbitrary,
and depending on the scientific question at hand, one might prefer to
apply other selection criteria. We have experimented with different
samples, and found most results to be remarkably robust, at least in a
qualitative way. In those cases where this is {\it not} the case, we
will mention so explicitly in the text.

\subsection{Segregation Properties and Indicators}
\label{sec:properties}

The main goal of this paper is to investigate how dark matter
subhaloes are segregated. Although segregation is most often discussed
in terms of radial segregation by mass (i.e., more massive objects are
located at smaller radii), in this paper we take a more comprehensive
approach, and examine the segregation of a wide variety of subhalo
properties; not just mass. We differentiate between {\it segregation
  indicators}, such as radius or binding energy, which express a
relation between the subhalo and its host halo, and {\it segregation
  properties}, such as mass or spin parameter, which are (mainly
internal) properties of the subhalo. The goal of this study is to
investigate which of the segregation properties are most segregated,
and how this segregation manifests itself as function of the different
segregation indicators.
\begin{table*}\label{tab:properties}
\caption{Segregation Properties}
\begin{center}
\begin{tabular}{ll}
\hline\hline
 $m$          & the present-day mass of the subhalo \\
 $m_{\rm acc}$ & the mass of the subhalo at accretion \\
 $m_{\rm peak}$ & the peak mass of the subhalo, defined as the maximum mass it ever had along its entire history. \\
 $m/m_{\rm acc}$ & the ratio of the present-day subhalo mass to that at accretion \\
 $V_{\rm max} $ & the maximum circular velocity of the present-day subhalo \\
 $V_{\rm acc}$  & the value of $V_{\rm max}$ of the subhalo at accretion \\
 $V_{\rm peak}$ & the peak value of $V_{\rm max}$ along the halo's history \\
 $V_{\rm max}/V_{\rm acc}$ & the ratio of present-day $V_{\rm max}$ to that at accretion \\
 $z_{\rm acc}$ & the subhalo's redshift of accretion \\
 $z_{\rm form}$ & the subhalo's formation redshift, defined as redshift at which main
progenitor reaches mass equal to $m_{\rm peak}/2$ \\
 $\lambda$ & the present-day spin parameter of the subhalo, as defined by Eq.~(\ref{spin}) \\
 $T/|U|$ & the virial parameter, with $T$ and $U$ the total kinetic and potential energy of the subhalo \\
\hline\hline
\end{tabular}
\end{center}
\medskip
\begin{minipage}{\hdsize}
\end{minipage}
\end{table*}

We consider the following segregation indicators:
\begin{itemize}
\item $r/r_{\rm vir}$ the three-dimensional distance between the
  center of the subhalo and the center of its host halo, normalized by
  the virial radius of the host halo.
\item $R/R_{\rm vir}$ the two-dimensional, projected distance between
  the center of the subhalo and the center of its host halo,
  normalized by the virial radius of the host halo. Throughout, we
  compute $R$ by simply projecting all haloes and subhaloes along the
  $z$-direction of the simulation box.
\item $E/|\Phi_0|$ the (specific) orbital energy of the subhalo,
  normalized by the value of the host halo's gravitational, central
  potential.
\end{itemize}
The specific orbital energy is computed as $E = v^2/2 + \Phi(r)$,
where
\begin{equation}\label{velocity}
v = {M_0 \over M_0 + m} \, \vert \vec{v}_{\rm host} - \vec{v}_{\rm sub} \vert
\end{equation}
is the speed of the subhalo with respect to the center of mass of the
host+subhalo system (here $\vec{v}_{\rm host}$ and $\vec{v}_{\rm sub}$
are the velocities of host halo and subhalo with respect to the
simulation box), and $\Phi(r)$ is the potential energy at the location
$r$ of the subhalo. We compute the latter using Eq.~(\ref{PhiNFW})
assuming that the host halo is a spherical NFW halo with a
concentration parameter, $c$, that we extract from the halo
catalog. Bound orbits have $-1 \leq E/|\Phi_0| < 0$, with more
negative values indicating a more bound orbit.  Clearly, $E/|\Phi_0|$
is the most physical segregation indicator, whereas $R/R_{\rm vir}$ is
the one that is most easily accessible observationally. We emphasize
that, when using $R/R_{\rm vir}$, we still only consider true
subhaloes that are located within the (3D) virial radius of the
host. The impact of interlopers is beyond the scope of this paper, but
will be addressed in detail in a follow-up paper in which we compare
our findings to observations (Lu et al., in preparation).

In addition to these three segregation indicators, we consider the
twelve segregation properties listed in Table~2. Throughout we
normalize the subhalo masses $m$, $m_{\rm acc}$ and $m_{\rm peak}$ by
the mass, $M_0$, of the present day host halo, and the subhalo
velocities $V_{\rm max}$, $V_{\rm acc}$, and $V_{\rm peak}$ by the
virial velocity, $V_{\rm vir,0}$, of the present-day host halo.  The
subhalo's accretion redshift, $z_{\rm acc}$, is defined as the {\it
  last} epoch at which the subhalo entered the virial radius of its
present-day host, while the formation redshift, $z_{\rm form}$, is
defined as the {\it first} epoch at which the subhalo's main
progenitor reaches a mass equal to, or larger than, $m_{\rm
  peak}/2$. Finally, regarding the virial ratio $T/|U|$, we emphasize
that we have ignored any external potential (e.g., such as that from
the host halo) when computing the potential energy, $U$, of the
subhaloes. If the subhalo is in isolation and in virial equilibrium
with negligble surface pressure, then we expect that $T/|U|=1/2$
(see e.g., \S~5.4.4 in Mo \etal 2010).
\begin{figure}
\centerline{\psfig{figure=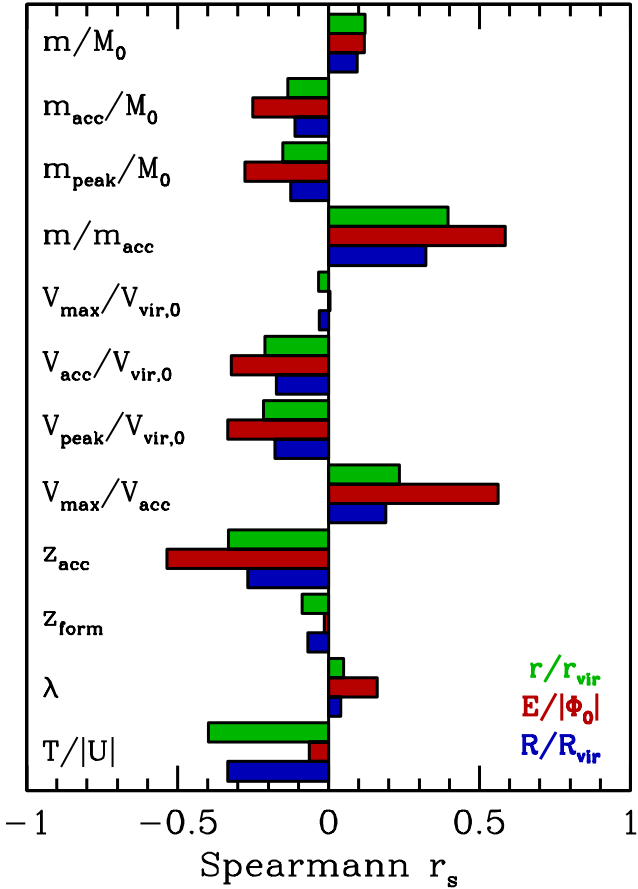,width=\hssize}}
\caption{Bar graph depicting the values of the Spearmann rank-order
  correlation coefficient, $r_\rms$, for the twelve subhalo properties
  (as indicated) with respect to $r/r_{\rm vir}$ (green bars),
  $E/|\Phi_0|$ (red bars) and $R/R_{\rm vir}$ (blue bars).}
\label{fig:bar}
\end{figure}

\subsection{Quantifying Segregation Strength}
\label{sec:strength}

For each combination of segregation property, $P$, and segregation
indicator, $S$, we compute the Spearmann rank-order correlation
coefficient, $r_\rms$, as an indicator of the segregation strength. We
have experimented with a number of alternative indicators, including
Kendall's tau and Pearson's linear correlation coefficient, which
always gave results in good, qualitative agreement\footnote{Typically,
  Pearson's linear correlation coefficient agrees with $r_\rms$ to
  better than $15\%$.}  with those based on $r_\rms$.  No segregation
corresponds to $r_\rms = 0$, while positive (negative) values of
$r_\rms$ indicate that $P$ is positively (negatively) correlated with
$S$ (i.e., we say that subhalo property $P$ reveals segregation with
respect to indicator $S$). In order to gauge the significance of a
non-zero $r_\rms$, we compute $r_\rms$ for 1000 Monte-Carlo samples in
which we reshuffle the rank-orders of $P$, while keeping the
rank-orders of $S$ fixed. By construction, these shuffled samples have
no segregation (i.e., no rank-order correlation), and the standard
deviation of $r_\rms$ among those 1000 values, $\sigma_r$, is
indicative of the variance in $r_\rms$ that one can obtain in the
absence of segregation. Since $r_\rms$ is non-parametric, $\sigma_{r}$
only depends on the number of subhaloes, and is independent of $P$ and
$S$. For the sample of subhaloes used here, we find that $\sigma_r =
0.0040$, indicating that a value of $|r_\rms| > 0.012$ corresponds to
a significant rank-order correlation between $P$ and $S$ at $99.9$\%
confidence.

In addition to $r_\rms$, which is non-parametric, we also quantify
the segregation strength using the parameter
\begin{equation} \label{Rparam}
\calR \equiv {\langle P \rangle_{0-25} \over \langle P \rangle_{75-100}}\,,
\end{equation}
where $\langle P \rangle_{0-25}$ and $\langle P \rangle_{75-100}$ are
the averages of $P_i$ for the quartiles of subhaloes with the lowest
and highest values of $S_i$, respectively. If $\calR > 1$ ($<1$) it
indicates that subhaloes in the lower quartile of $S_i$ have a
property $P$ that is $\calR$ times larger (smaller) than for the
subhaloes in the upper quartile of $S_i$. The advantage of $\calR$ is
that it is easier to interpret, whereas $r_\rms$ better expresses the
statistical significance of the segregation. We caution, though, that
$r_\rms$ is only a meaningful segregation strength indicator if the
correlation between $P$ and $S$ is monotonic. As we show below, this
is not always the case.


\section{Results}
\label{sec:res}

As indicated in \S\ref{sec:properties} we have a total of 12 subhalo
properties, $P$, and three segregation indicators, $S$. The values of
$r_\rms$ and $\calR$, for all 36 combinations of segregation indicator
and property, are listed in Table~3.  Note that {\it all} twelve
subhalo properties are significantly segregated (in that $|r_\rms| >
0.012$) with respect to {\it each} of the three segregation
indicators, but for one exception: $\Vmax$, the present-day maximum
circular velocity of subhaloes, normalized by the virial velocity of
the host halo, is the only subhalo property that is consistent with
having no significant correlation with orbital binding energy.

As a visualization of these results, the bar graph in
Fig.~\ref{fig:bar} shows the Spearmann rank-order correlation
coefficients for all twelve subhalo properties, with green, red
and blue bars indicating the value of $r_\rms$ with respect to
$r/r_{\rm vir}$, $E/|\Phi_0|$, and $R/R_{\rm vir}$, respectively.
A few trends are immediately evident. As expected, segregation with
respect to the {\it projected} radius $R/R_{\rm vir}$ is always weaker
than with respect to the 3D radius, but only by a little bit. In
particular, the value of $|r_\rms|$ in projection is typically $20\%$
smaller than that in 3D, while the difference is even smaller for the
$\calR$ parameter. In virtually all cases, the absolute value of
$r_\rms$ is largest when using $E/|\Phi_0|$ as indicator, which is
what one expects if the segregation is physical. Two notable outliers
in this respect are $T/|U|$ and $z_{\rm form}$. The virial ratio in
particular shows strong segregation (large values of $|r_\rms|$) when
using $r/r_{\rm vir}$ or $R/R_{\rm vir}$ as indicators, but has one of
the smallest values for $|r_\rms|$ when using $E/|\Phi_0|$ as
segregation indicator. As we discuss in \S\ref{sec:virial} below, this
is a consequence of impulsive heating at peri-centric passage.
\begin{table}\label{tab:spear}
\caption{Segregation Strength Indicators}
\begin{center}
\begin{tabular}{llrrr}
\hline\hline
 Seg. Prop.     & Stat & $r/r_{\rm vir}$ & $E/|\Phi_0|$ & $R/R_{\rm vir}$ \\
 (1) & (2) & (3) & (4) & (5) \\
\hline
$\msub$                  & $r_\rms$ & 0.120 & 0.118 & 0.095 \\
                          & $\calR$ &  1.058 & 1.416 & 1.058 \\
$\macc$                  & $r_\rms$ & -0.136 & -0.251 & -0.112 \\
                          & $\calR$  & 1.644 & 2.643 & 1.537 \\
$\mpeak$                 & $r_\rms$ & -0.151 & -0.277 & -0.125 \\
                          & $\calR$ & 1.687 & 2.821 & 1.573 \\
$\mrat$                  & $r_\rms$ &  0.397 &  0.585 &  0.322 \\
                          & $\calR$ & 0.609 & 0.479 & 0.668 \\
$V_{\rm max}/V_{\rm vir,0}$   & $r_\rms$ & -0.034 & 0.005 & -0.031 \\
                          & $\calR$ & 1.052 & 1.057 & 1.045 \\
$V_{\rm acc}/V_{\rm vir,0}$   & $r_\rms$ & -0.211 & -0.321 & -0.173 \\
                          & $\calR$ & 1.169 & 1.301 & 1.141 \\
$V_{\rm peak}/V_{\rm vir,0}$  & $r_\rms$ & -0.215 & -0.335 & -0.177 \\
                          & $\calR$ &  1.171 & 1.312 & 1.144 \\
$V_{\rm max}/V_{\rm acc}$     & $r_\rms$ &  0.235 &  0.561 &  0.189 \\
                          & $\calR$ & 0.904 & 0.810 & 0.920 \\
$z_{\rm acc} $              & $r_\rms$ & -0.332 & -0.536 & -0.268 \\
                          & $\calR$ & 2.119 & 3.131 & 1.862 \\
$z_{\rm form} $              & $r_\rms$ & -0.087 & -0.014 & -0.069 \\
                          & $\calR$ & 1.104 & 1.021 & 1.082 \\
$\lambda$                  & $r_\rms$ &  0.050 &  0.161 &  0.040 \\
                          & $\calR$ & 0.927 & 0.798 & 0.938 \\
$T/|U|$                    & $r_\rms$ & -0.399 & -0.063 & -0.335 \\
                          & $\calR$ & 1.207 & 1.043 & 1.174 \\
\hline\hline
\end{tabular}
\end{center}
\medskip
\begin{minipage}{\hssize}
For each segregation property (column 1), this table lists the values
of the Spearmann rank-order correlation coefficient, $r_\rms$, and the
ratio $\calR$ defined by Eq.~(\ref{Rparam}), for the segregation
indicators $r/r_{\rm vir}$ (column 3), $E/|\Phi_0|$ (column 4), and
$R/R_{\rm vir}$ (column 5).
\end{minipage}
\end{table}

Next we proceed as follows.  For a given set of $S$ and $P$, we first
rank-order the subhaloes by their segregation indicator $S$. Next we
split the rank-ordered sample in 40 equal-number bins, and compute,
for each bin, the averages of $S$ and $P$. The results for 8 different
segregation properties\footnote{For the sake of brevity, we only show
  results for 8 out of 12 properties.}, $P$, are indicated as open
circles in Figs.~\ref{fig:seg1} and~\ref{fig:seg2}. Different columns
(rows) correspond to different segregation indicators (properties),
and the corresponding value for $r_\rms$ is indicated in the
lower-left corner of each panel. Error bars indicate the range of
segregation indicator for each bin. Different colors correspond to the
three different simulation boxes, as indicated, and are typically in
excellent mutual agreement. Plotting the medians of $P$, rather than
the averages, yields plots that are virtually indistinguishable.  In
what follows we discuss some individual results in turn.

\subsection{Masses and Circular Velocities}
\label{sec:massresults}

\begin{figure*}
\centerline{\psfig{figure=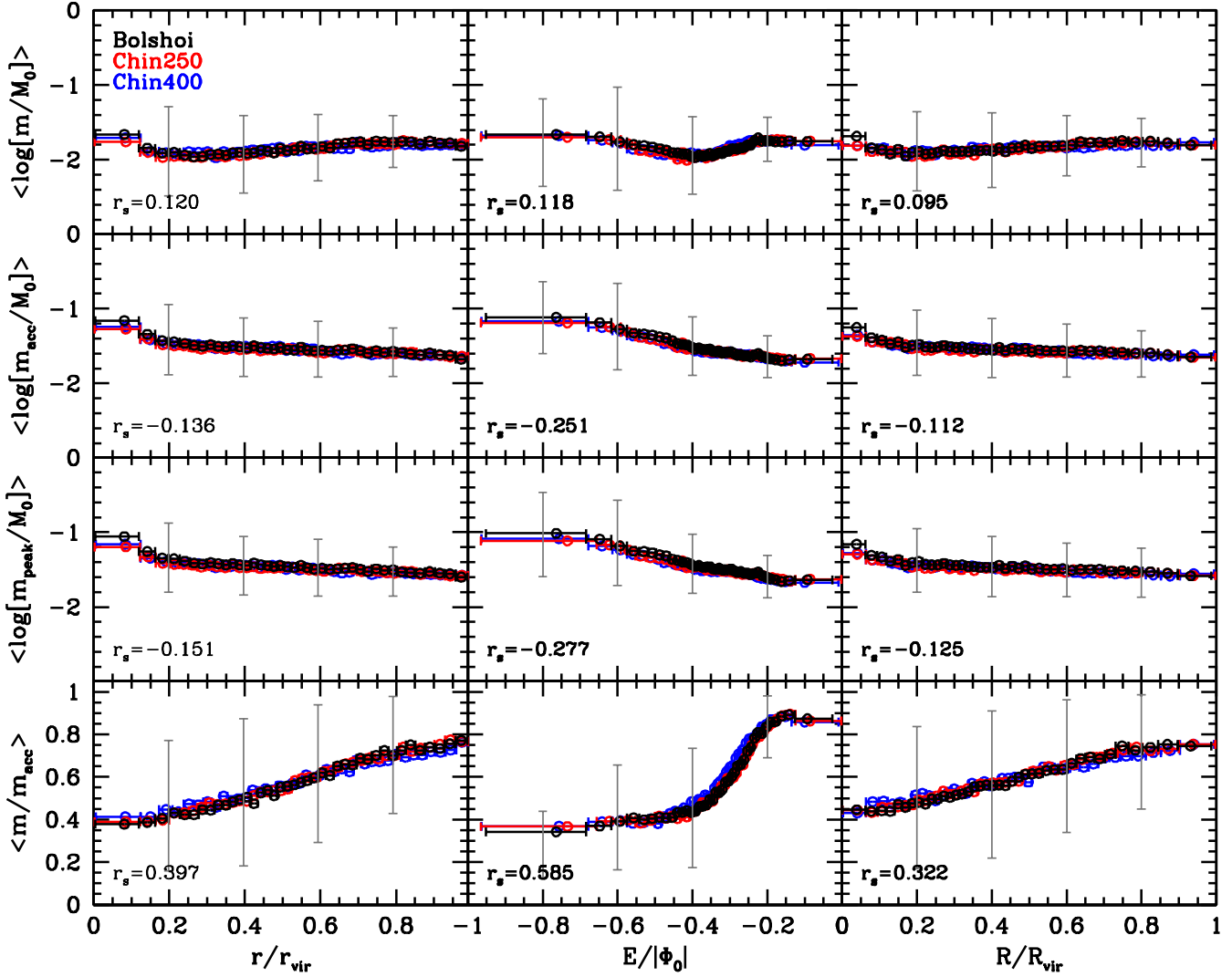,width=\hdsize}}
\caption{Each panel plots an average segregation property along the
  y-axis as function of a segregation indicator along the x-axis. The
  corresponding Spearmann rank-order correlation coefficient is
  indicated in the lower-left corner of each panel, while different
  colors correspond to different simulation boxes, as indicated. From
  top to bottom the segregation properties are the present-day subhalo
  mass, $\log[\msub]$, the subhalo mass at accretion, $\log[\macc]$,
  the peak subhalo mass $\log[\mpeak]$ (all normalized by the
  present-day host halo mass, $M_0$), and the mass ratio $m/m_{\rm
    acc}$. From left-to-right, the segregation indicators are the
  halo-centric distance, $r/r_{\rm vir}$, the orbital energy,
  $E/|\Phi_0|$, and the projected halo-centric distance, $R/R_{\rm
    vir}$. Horizontal errorbars indicate the width of each of the 40
  bins used, which are chosen to have an equal number of subhaloes in
  them. Vertical errorbars indicate the 16 to 84 percentile range of
  the segregation properties; to avoid clutter, these are only plotted
  for four bins, and only for the Bolshoi simulation.}
\label{fig:seg1}
\end{figure*}

\begin{figure*}
\centerline{\psfig{figure=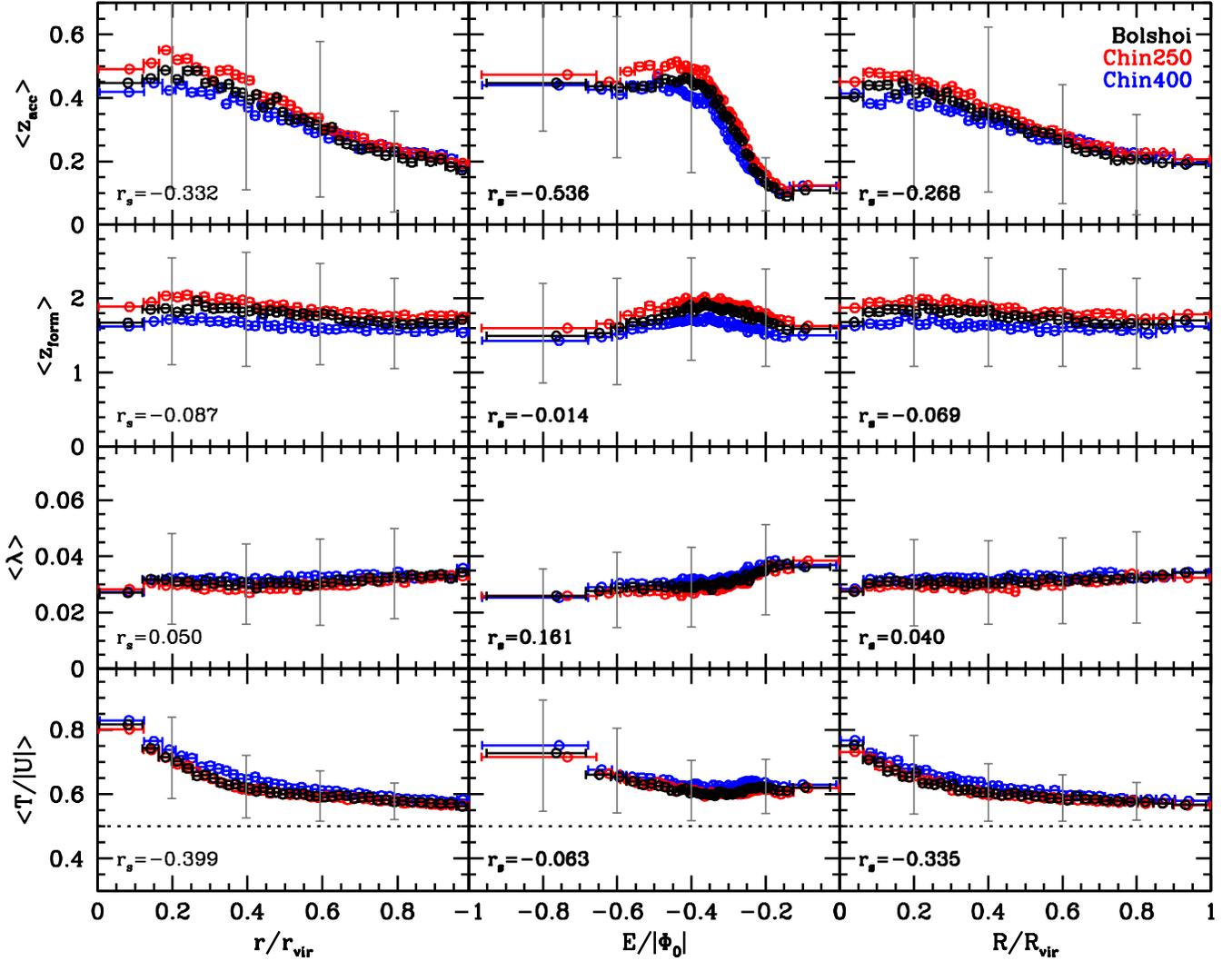,width=\hdsize}}
\caption{Same as Fig.~\ref{fig:seg1}, but for the subhalo accretion
  redshift $z_{\rm acc}$ (upper panels), the subhalo formation
  redshift $z_{\rm form}$ (second row of panels), the subhalo spin
  parameter $\lambda$ (third row of panels), and the ratio of kinetic
  to potential energy, $T/|U|$ (lower panels). The dotted, horizonal
  lines in the lower panels correspond to virial equilibrium (i.e.,
  $T/|U|=1/2$).}
\label{fig:seg2}
\end{figure*}

The first subhalo property we discuss is the mass ratio $\mrat$, whose
correlation with respect to the three segregation indicators is
indicated in the lower panels of Fig.~\ref{fig:seg1}.  Clearly,
$\mrat$ is extremely strongly segregated, in the sense that subhaloes
that have lost a larger fraction of their mass at accretion are on
more bound orbits. This was first hinted at in the studies by Gao
\etal (2004) and Nagai \& Kravtsov (2005). Based on the $\calR$
statistic, the most bound quartile has, on average, lost more than two
times as much mass than the least bound quartile, and with $r_\rms =
0.585$, the mass ratio $\mrat$ is the subhalo property that reveals
the strongest segregation among all 12 properties examined in this
study. As we demonstrate in \S\ref{sec:mrat}, this has its origin in
the inside-out assembly of haloes combined with the impact of tidal
stripping.


The second row of panels in Fig.~\ref{fig:seg1} shows that subhaloes
are also strongly segregated by $\macc$, to the extent that subhaloes
that were more massive at accretion are on more bound orbits, and
located at smaller halo-centric distances (see also Contini \etal
2012). The most bound quartile of subhaloes has accretion masses that
are $2.6$ times larger than for the least bound quartile. Results for
$\mpeak$, shown in the third row of panels, are very similar to those
for $\macc$.  Based on the corresponding $r_\rms$ and $\calR$ values,
$m_{\rm peak}$ is slightly more segregated than $m_{\rm acc}$, but
only by a small amount. As we will demonstrate in \S\ref{sec:macc},
the segregation of $m_{\rm acc}$ and $m_{\rm peak}$ is a consequence
of dynamical friction during the first orbital period, and is weakened
by tidal disruption and by the fact that more massive subhaloes are
typically accreted later.

The upper panels of Fig~\ref{fig:seg1} show the results for the
present-day mass, $m$, normalized by that of its host halo, $M_0$. As
it turns out, for this property it is difficult to discern a unique
sign of segregation.  Based on the values of $r_\rms$, which are
positive for all three segregation indicators, more massive subhaloes
are {\it less} bound and located at ${\it larger}$ halo-centric
radii. Although in qualitative agreement with results from several
other simulations (De Lucia \etal 2004; Reed \etal 2005; Angulo \etal
2009; Contini \etal 2012), this is opposite to naive expectations that
more massive subhaloes should be more bound due to dynamical friction.
As we will see below, this simple prediction is complicated by the
fact that subhaloes experience appreciable amounts of mass loss. More
confusing yet is the fact that the $\calR$ values for $\msub$ are
larger than unity, indicating that the lower quartile of the $r/r_{\rm
  vir}$ and $E/|\Phi_0|$ distributions are actually {\it more massive}
than the corresponding upper quartiles: this is opposite to the trend
one seems to infer from the Spearmann rank-order correlation
coefficient. This apparent contradiction is explained by the fact that
the trends of $\langle \log[\msub]\rangle$ with $r/r_{\rm vir}$,
$E/|\Phi_0|$ and $R/R_{\rm vir}$ are not monotonic.

In general, we find the results for the present-day halo mass to be
extremely sensitive to sample selection. For example, if we were to
include subhaloes with $-3 \leq \log[\macc] < -2$ (i.e., we add
subhaloes located in the pink-shaded, triangular region of the
right-hand panel of Fig.~\ref{fig:sample}), then the $r_\rms$ values
for all three segregation indicators become negative\footnote{In the
  case of $E/|\Phi_0|$, $r_\rms$ changes from $0.118$ to $-0.095$.},
and the equivalent of the upper panels of Fig.~\ref{fig:seg1} now
shows a monotonic decline of $\langle \log[\msub]\rangle$ with
increasing $r$, $E$ and $R$. Hence, {\it depending on the exact
  selection criteria used, subhaloes can reveal very different levels,
  and signs, of mass segregation.} The reason is, as we have already
seen, that subhaloes are extremely strongly segregated in $\mrat$. As
a consequence, if different sample selections make slightly different
cuts in the distribution of $\mrat$, this can have strong impact on
the results for $\msub$ (and also, albeit to a lesser extent, for
$\macc$). We also believe this to be the reason for a lack of
consensus regarding the sign and strength of subhalo mass segregation
in previous studies (see discussion in \S\ref{sec:intro}).
Appendix~\ref{App:sample} presents a more detailed, complementary
description of the segregation of $\msub$ and $\macc$, together with a
demonstration of the impact of sample selection.

Although not shown in Fig.~\ref{fig:seg1}, the results for $V_{\rm
  max}$, $V_{\rm acc}$, $V_{\rm peak}$ and $V_{\rm max}/V_{\rm acc}$
are qualitatively very similar to those for their corresponding mass
indicators, $m$, $m_{\rm acc}$, $m_{\rm peak}$, and $\mrat$.
Segregation is a bit stronger for $\Vacc$ and $\Vpeak$ than for their
respective masses. However, $\Vmax$ is the property with the overall
weakest segregation among all 12 properties studies here. Note,
though, that similar to $m/M_0$, the segregation strength (and sign)
of $\Vmax$ depends strongly on sample selection. We emphasize that a
comparison of segregation strengths should be done based on the
non-parametric $r_\rms$, and not $\calR$. In fact, the $\calR$ values
for the maximum circular velocities are closer to unity, in line with
the expectations based on the fact that to reasonable approximation
dark matter haloes have $V_{\rm max} \propto M^{1/3}$.

\subsection{Accretion Redshift, Formation Redshift, Spin Parameter and Virial Ratio}
\label{sec:other}

The upper panels of Fig.~\ref{fig:seg2} show that subhaloes are also
extremely strongly segregated by their redshift of accretion, $z_{\rm
  acc}$. Subhaloes that were accreted earlier are more bound, and
located at smaller halo-centric radii, in qualitative agreement with
the results of Gao \etal (2004), Faltenbacher \& Diemand (2006), and
Contini \etal (2012). The fact that there is a small offset between
the Chin250 and Chin400 results reflects the differences in host halo
mass; since we only selected subhaloes in host haloes with at least $5
\times 10^4$ particles, the average host halo mass in the Chin400
simulation is somewhat larger than in the Chin250 (and Bolshoi)
simulations. Since more massive haloes assemble later (e.g., van den
Bosch 2002), the average accretion redshift of subhaloes in Chin400 is
smaller than for Chin250. The small difference between the Chin250 and
Bolshoi simulations, which have (almost) the same mass resolution,
arises from the fact that Chin250 used fewer simulation outputs,
causing a reduced temporal resolution of the halo merger trees, which
biases the accretion redshifts high. We find that the most bound
quartile of subhaloes typically was accreted 2.5 to 3.0 Gyr earlier
than the least bound quartile (for the range of host halo masses
examined here).  As we will see in \S\ref{sec:zacc}, this segregation
by accretion redshift is mainly a reflection of the inside-out
assembly of dark matter haloes.

The second row of panels of Fig.~\ref{fig:seg2} shows that the average
formation redshift for the subhaloes in our sample is $\sim 1.8$, and
that $z_{\rm form}$ is weakly segregated in the sense that subhaloes
that formed earlier are located further in, and on more bound orbits.
The segregation is weak though, especially with respect to
$E/|\Phi_0|$ for which $r_\rms=-0.014$ (one of the least significant
Spearmann correlation coefficient of all 36 listed in Table~3). Note
also that, similar to present-day mass, the correlation between
$z_{\rm form}$ and $E/|\Phi_0|$ is non-monotonic.

The third row of panels of Fig.~\ref{fig:seg2} shows that the spin
parameter of dark matter subhaloes (as defined by Eq.~[\ref{spin}]) is
only weakly segregated, in the sense that subhaloes on more bound
orbits have somewhat lower $\lambda$. This is consistent with the
findings of Reed \etal (2005) and Onions \etal (2013).  As indicated
by the $\calR$ value, the most bound quartile has a spin parameter
than is $\sim 20\%$ smaller than for the least bound quartile. As
discussed in \S\ref{sec:spin}, this segregation of spin parameters is
a manifestation of the fact that more bound subhaloes have experienced
more mass stripping (cf. lower panels of Fig.~\ref{fig:seg1}), and
stronger tidal torques from their host.

Finally, the lower panels of Fig.~\ref{fig:seg2} show the results for
$T/|U|$. These show that the virial ratio is strongly segregated with
respect to radius, in the sense that subhaloes at smaller halo-centric 
distances have $T/|U|$ ratios that deviate more from $1/2$, the value
corresponding to virial equilibrium and indicated by horizontal,
dotted lines. Although difficult to judge from Fig.~\ref{fig:seg2},
the $r_\rms$ and $\calR$ values listed in Table~3 indicate that
whereas $T/|U|$ is strongly segregated with respect to $r/r_{\rm vir}$
and $R/R_{\rm vir}$, this is not the case with respect to
$E/|\Phi_0|$. In fact, with $r_\rms = -0.063$ the rank-order
correlation between the virial ratio and orbital energy is one of the
weakest correlations among the ensemble of 36 listed.


\section{The origin of subhalo segregation}
\label{sec:orig}

In the previous section we described how each of the twelve subhalo
properties is segregated with respect to each of the three segregation
indicators.  We now turn to address the origin of this segregation.
We will do so roughly in the reverse order in which we presented the
results in the previous section, starting with the segregation of the
virial ratio, $T/|U|$.  We will not discuss the velocity-based
properties $\Vmax$, $\Vacc$, $\Vpeak$ and $\Vrat$, but note that their
segregation origin is akin to that of their corresponding mass-based
property.
\begin{figure*}
\centerline{\psfig{figure=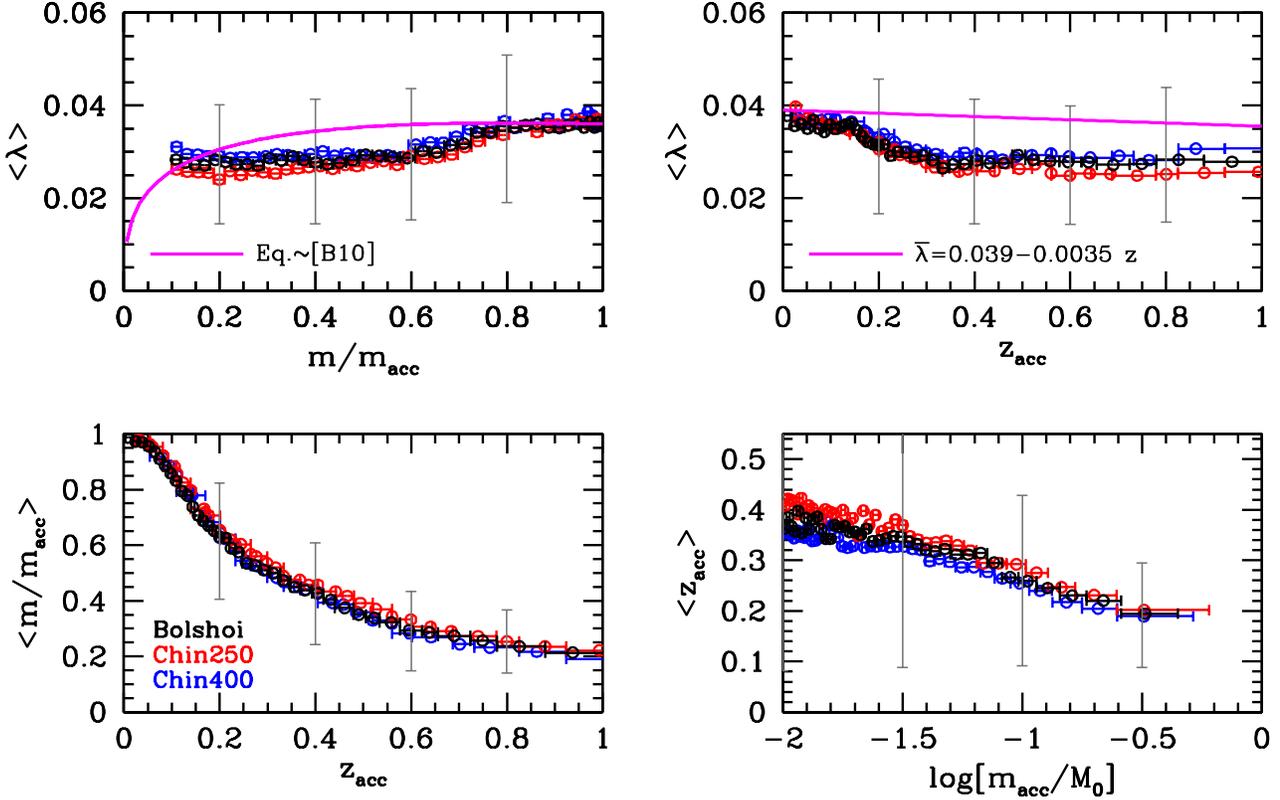,width=0.95\hdsize}}
\caption{Correlations between subhalo properties. Clockwise from the
  upper left, the panels show the average spin parameter, $\lambda$,
  as function of $m/m_{\rm acc}$ and $z_{\rm acc}$, the average mass
  ratio $m/m_{\rm acc}$ as function of $z_{\rm acc}$, and the average
  accretion redshift, $z_{\rm acc}$, as function of $\log[m_{\rm
      acc}/M_0]$. As in Figs.~\ref{fig:seg1} and~\ref{fig:seg2},
  different colors correspond to different simulations, as indicated,
  horizontal errorbars indicate bin width, and vertical errorbars
  indicate the 16-84 percentiles. Note that (ii) subhaloes that have
  lost a large fraction of their mass at accretion have lower spin,
  (ii) subhaloes that were accreted more recently have larger spin,
  (iii) subhaloes that are accreted later have lost a smaller fraction
  of their mass to tidal stripping, and (iv) subhaloes with a larger
  mass at accretion are accreted later. The thick, magenta line in the
  upper-left panel corresponds to a simple prediction based on
  outside-in stripping (see \S\ref{sec:spin} and
  Appendix~\ref{App:spin}), while the thick, magenta line in the
  upper-right panel indicates the redshift dependence of the average
  halo spin parameter of Hetznecker \& Burkert (2006). }
\label{fig:propcorr}
\end{figure*}

\subsection{Segregation of Virial Ratio}
\label{sec:virial}

As shown in \S\ref{sec:other}, the segregation of the virial ratio
$T/|U|$ is strong with respect to $r/r_{\rm vir}$, but almost absent
with respect to the orbital energy, $E/|\Phi_0|$. This suggests that
strong enhancements of $T/|U|$, reflecting strong deviations from
virial equilibrium, are not a characteristic of certain orbits, but
rather of certain orbital phases, and indicates that the radial
segregation of $T/|U|$ is due to impulsive heating.  During its
peri-centric passage, the subhalo experiences a rapidly varying
external potential (due to the host halo).  According to the impulse
approximation, this increases the kinetic energy of the constituent
particles, causing a boost in $T/|U|$ that brings the subhalo out of
virial equilibrium (see e.g., Mo, van den Bosch \& White 2010).
Subsequently, as the subhalo travels back out to apo-center,
re-virialization drives $T/|U|$ of the subhalo back to $1/2$. Hence,
subhaloes predominantly have an enhanced value of $T/|U|$ close to
peri-center, and not along their entire orbit (see also
\S\ref{sec:zacc} below). This explains why $T/|U|$ is (strongly)
segregated with respect to radius, but only weakly with respect to
orbital energy. The weak segregation with respect to $E/|\Phi_0|$
arises because more bound orbits have smaller peri-centric distances,
and therefore experience stronger tidal shocking, resulting in a
larger {\it average} $T/|U|$ along their orbit.

As is evident from the lower panels of Fig.~\ref{fig:seg2}, the
average $T/|U|$ is always well in excess of the virial value, $1/2$,
indicating that the average subhalo is never really in virial
equilibrium. This is expected given that the internal dynamical time
of a subhalo is basically the same as that of its host halo.  Given
that it takes at least a dynamical time to re-adjust virial
equilibrium after an impulsive shock, and given that subhaloes
typically experience one impulsive shock per orbit (during
peri-centric passage), it should not come as a surprise that $\langle
T/|U|\rangle > 0.5$. In addition, we emphasize that the surface
pressure on subhaloes can also be substantial, which may also 
contribute to boosting the virial ratio above one half.

\subsection{Segregation of Spin Parameter}
\label{sec:spin}

As shown in Fig.~\ref{fig:seg2}, subhaloes with smaller spin
parameters are segregated towards the center of their host halo, in
good agreement with the findings of Reed \etal (2005) and Onions \etal
(2013). Both Reed \etal and Onions \etal argue that high angular
momentum material is more vulnerable to stripping than low angular
momentum, so that mass stripping will tend to lower the spin
parameter. In this picture the segregation of spin parameter is merely
an outcome of the segregation of $m/m_{\rm acc}$.

This idea, that mass stripping lowers the spin parameter, is
qualitatively consistent with the notion that angular momentum is less
centrally concentrated within haloes than mass (e.g., Navarro \&
Steinmetz 1997; Bullock \etal 2001; Dutton \& van den Bosch 2012), and
is supported by the upper-left panel of Fig.~\ref{fig:propcorr}, which
shows that subhaloes that have lost a larger fracion of their mass
have lower spin parameters. To test this idea in detail, we have
computed the average spin parameter profile, $\lambda(<r)$, defined as
the spin parameter of the material interior to radius $r$. We assume
that halos follow a NFW density distribution with concentration
parameter $c=15$, and that their specific angular momentum at radius
$r$ follows $j(r) \propto [M(<r)]^{1.3}$, which is a good description
of the average specific angular momentum profile of dark matter haloes
(Bullock \etal 2001).  $M(<r)$ is the mass enclosed within radius
$r$. Next we use this spin parameter profile to predict the relation
between $\lambda$ and $\mrat$ by assuming that the spin parameter of a
subhalo is equal to its original spin (prior to mass stripping) inside
the radius that encloses the mass $m$ (see Appendix~\ref{App:spin} for
details). In other words, it is assumed that stripping proceeds in an
outside-in fashion, much like peeling the layers of an onion, and that
the angular momentum of the non-stripped material is unaffected.  The
solid, magenta line in the upper-left panel of Fig.~\ref{fig:propcorr}
shows the resulting prediction for the relation between $\lambda$ and
$\mrat$, where we have normalized $\lambda$ to fit the simulation data
at $\mrat = 1$. This simple `onion model' predicts that $\lambda$
increase with $\mrat$, in qualitative agreement with the simulation
results\footnote{This is opposite to the average spin profile derived
  for subhaloes by Onions \etal (2013); However, as we show in
  Appendix~\ref{App:spin}, the Onions \etal results are an artefact of
  poorly resolved haloes.}. However, in detail the prediction is
clearly a poor description of the data.

Consequently, we do not think that this `onion' model is the complete
picture. In fact, there are good reasons to expect it to be
incomplete. After all, whenever tidal forces on a subhalo are large
enough to strip a significant fraction of its mass, clearly the tidal
forces are also strong enough to torque the remaining core. Hence, the
spin parameter of subhaloes is likely to be directly influenced, and
modified, by the tidal field of its host.  Indeed, several studies
have shown that subhaloes tend to be radially aligned, in that their
major axis points in the direction of the center of the host halo
(e.g., Kuhlen \etal 2007; Pereira, Bryan \& Gill 2008; Faltenbacher
\etal 2008). This alignment is largely preserved as the subhalo
undergoes its peri-centric passage, indicating that it indeed
experiences a torque from the host. We leave it for a future study
to investigate how exactly such torques modify the spin parameters of
subhaloes. At this stage, we merely point out that is likely an
important, additional mechnanism for understanding the spin parameters
of dark matter subhaloes, including their segregation.
\begin{figure}
\centerline{\psfig{figure=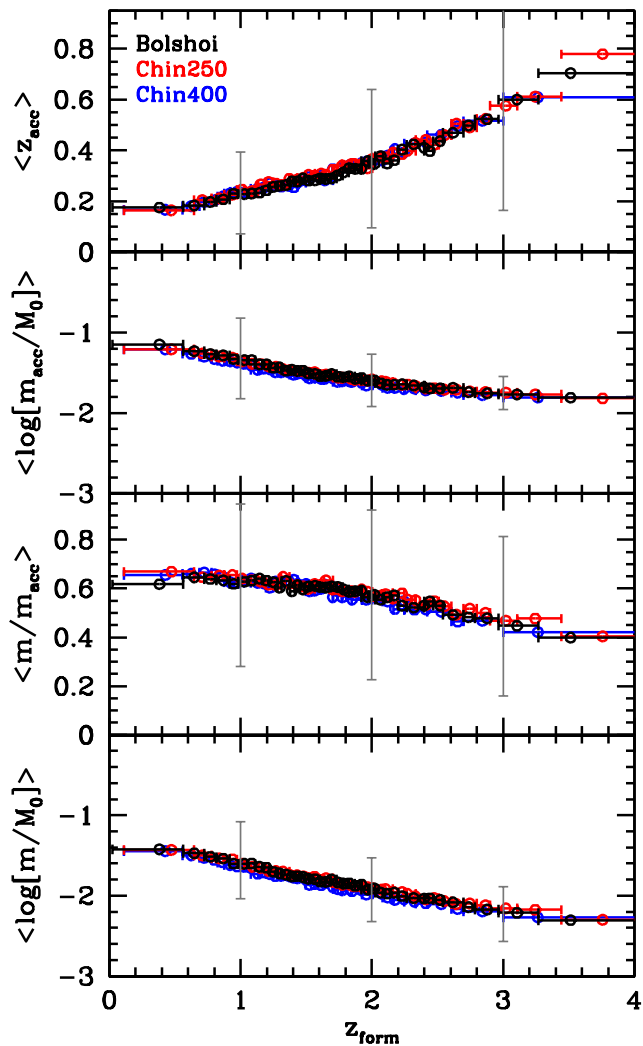,width=\hssize}}
\caption{Correlations between various subhalo properties and the
  subhalo's formation redshift. From top to bottom, the subhalo
  properties plotted are the accretion redshift, the mass at
  accretion, the mass ratio $m/m_{\rm acc}$, and the present-day
  subhalo mass. The various trends are discussed in detail in
  \S\ref{sec:zform}.  Color coding and \st{horizontal} errorbars are as in
  the previous figures.}
\label{fig:zform}
\end{figure}

Finally, we briefly mention another effect that might contribute to
the segregation of subhalo spin.  As shown by Hetznecker \& Burkert
(2006), the average spin parameter of dark matter haloes evolves
with redshift as
\begin{equation}\label{spinevolve}
\bar{\lambda}(z) \simeq 0.039 - 0.0035 \, z\,.
\end{equation}
If we postulate that this redshift evolution is arrested when a halo
becomes a subhalo, then subhaloes that were accreted earlier would
have a smaller spin parameter on average (at least at infall).  Since
haloes that accreted earlier are segregated towards the center, this
could potentially explain the observed segregation of subhalo spin
parameter. The upper-right panel of Fig.~\ref{fig:propcorr} shows that
indeed subhaloes with a lower spin parameter were accreted at higher
redshifts. However, the trend is inconsistent with the expectations
from Eq.~(\ref{spinevolve}), indicated by the thick, magenta line. The
redshift evolution of $\bar{\lambda}$ is simply too weak to have an
impact on the subhaloes, the majority of which were accreted
relatively recently ($z_{\rm acc} < 0.5$).  It is interesting, though,
that most of the correlation between $\lambda$ and $z_{\rm acc}$ seems
confined to the range $0.1 \lta z_{\rm acc} \lta 0.3$. As we will see
below, subhaloes accreted in this interval are between their first
peri- and first apo-centric passages, and the data seems to suggest
that this period is critical for regulating the subhalo spin
parameter.

\subsection{Segregation of Formation Redshift}
\label{sec:zform}

As we have shown, subhaloes are weakly segregated in their formation
redshift, with those that form earlier located closer to the center of
their host halo. As we demonstrate below, the origin of this
segregation is indirect, and arises from the fact that the formation
redshift of subhaloes is correlated with a number of other subhalo
properties, each of which is segregated.  Fig.~\ref{fig:zform} shows
how subhaloes with a higher formation redshift (i) are accreted
earlier, (ii) are less massive at accretion, (iii) have lost more mass
since accretion, and (iv) have a smaller present-day mass. Results (i)
and (ii) are manifestations of hierarchical structure
formation. Result (iii) is a natural outcome of (i), when one takes
into account that subhaloes that were accreted earlier have been
exposed to mass stripping for a longer period of time (see
\S\ref{sec:mrat}). And finally, result (iv) is a consequence of (ii)
and (iii) combined. Note that the correlation between $z_{\rm form}$
and $z_{\rm acc}$ has previously been pointed out in Hearin \etal
(2014).

In terms of segregation, result (i) would imply that subhaloes that
form earlier are segregated towards the center of their host halo.
After all, subhaloes with larger $z_{\rm acc}$ are on more bound
orbits.  This is qualitatively consistent with the fact that the
$r_\rms$ values for $z_{\rm form}$ are negative. However, this is
counter-balanced by result (ii), and the fact that subhaloes that are
less massive at accretion are segregated towards the outskirts of
their host halo. These two competing effects result in a final
segregation for $z_{\rm form}$ that is weak. It also explains why the
correlation between $z_{\rm form}$ and $E/|\Phi_0|$ is non-monotonic
(cf. Fig.~\ref{fig:seg2}).

\subsection{Segregation of Accretion Redshift}
\label{sec:zacc}

\begin{figure}
\centerline{\psfig{figure=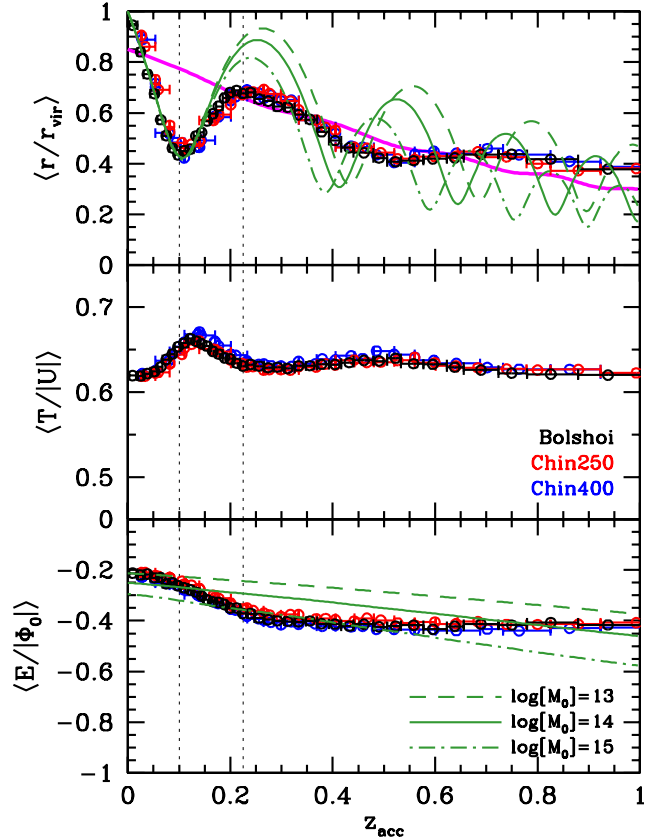,width=\hssize}}
\caption{As a function of subhalo accretion redshift, the various
  panels show, from top to bottom, the average halo-centric radius,
  $r/r_{\rm vir}$, the average virial ratio $T/|U|$, and the average
  orbital energy, $E/|\Phi_0|$. The green lines are predictions from
  the `adiabatic evolution model' described in the text, for three
  different host halo masses, $M_0$, as indicated in the lower
  panel. The thick, magenta curve is the prediction from the
  same model, but for subhaloes which at accretion start out on a
  circular orbit. The vertical dotted lines indicate the accretion
  redshifts of subhaloes that are currently (at $z=0$) experiencing
  their first peri- and apo-centric passage. Color coding and
  horizontal errorbars are as in the previous figures.}
\label{fig:orbit}
\end{figure}

Subhaloes are strongly segregated in their accretion redshift, with
early-accreted subhaloes residing on more bound orbits.  This has its
origin in the inside-out assembly of dark matter haloes.
Fig.~\ref{fig:orbit} plots the average, normalized radius, $\langle
r/r_{\rm vir}\rangle$, the average virial ratio, $\langle
T/|U|\rangle$, and the average orbital energy, $\langle
E/|\Phi_0|\rangle$, as functions of $z_{\rm acc}$. Note how the
average halo-centric distance, which is averaged over tens of
thousands of subhaloes, clearly reveals the characteristics of an
orbit, at least for $z_{\rm acc} \lta 0.5$.  Subhaloes accreted at
$z_{\rm acc} \sim 0.1$ have just reached their first peri-center,
while subhaloes that are at their first apo-centric passage after
accretion typically where accreted around $z_{\rm acc} \sim
0.25$. Note how the middle panel shows that the average $T/|U|$ of
subhaloes experiences a temporary boost shortly after first
peri-centric passage, in accordance with the impulsive heating picture
sketched above in \S\ref{sec:virial}.

As the lower panel shows, subhaloes accreted earlier are more bound.
This could have two explanations. Either this could simply reflect
that at earlier times the main progenitor of the host was smaller, so
that accretion `naturally' occurred at smaller halo-centric distances,
or it could be that subhaloes that were accreted earlier were exposed
to dynamical friction for a longer period, resulting in a more
negative binding energy today. We now proceed to demonstrate that the
former dominates.

Let $M(z)$, $r_{\rm vir}(z)$ and $c(z)$ be the virial mass, virial
radius and concentration of the main progenitor of a host halo of
present-day mass $M_0$. If we assume that this main progenitor can, at
any redshift, be described by an NFW profile, then the corresponding
time-varying gravitational potential is
\begin{equation}\label{pottvar}
\Phi(r,z) = -{G M(z) \over r_{\rm vir}(z)} \,
{\ln[1+x\, c(z)]\over x \, f[c(z)]}\,,
\end{equation}
where $x = r/r_{\rm vir}(z)$ and $f(x)$ is given by Eq.~(\ref{fx}). If
we assume that the time-variability of $\Phi(r,z)$ is sufficiently
slow, so that the system responds adiabatically, then, given an
initial (at accretion) orbital energy, $E_{\rm orb}$, and angular
momentum, $L_{\rm orb}$, and ignoring dynamical friction, we can
predict the current (i.e., at $z=0$) location and orbital energy of a
subhalo accreted at $z_{\rm acc}$, by simple integration of its orbit
in the time-varying potential $\Phi(r,z)$. We refer to this as our
`adiabatic evolution model'. In order to specify the initial orbit at
accretion, we introduce the orbital parameters $\alpha$ and
$\eta$, defined by
\begin{equation}\label{initial}
E_{\rm orb} = E_{\rm circ}(\alpha \, r_{\rm acc}) \,,\;\;\;\;\;\;\;\;\;\;
L_{\rm orb} = \eta \, L_{\rm circ}(\alpha \, r_{\rm acc})\,.
\end{equation}
Here $r_{\rm acc} \equiv r_{\rm vir}(z_{\rm acc})$ is the virial
radius of the main progenitor at the accretion redshift, and $E_{\rm
  circ}(r)$ and $L_{\rm circ}(r)$ are the orbital energy and angular
momentum of a circular orbit of radius $r$. Thus defined, $\eta = 0$
and $1$ correspond to purely radial and circular orbits, respectively,
while $\alpha =1$ corresponds to the orbital energy of a circular
orbit with radius $r = r_{\rm acc}$.

We now proceed as follows. For a given $M_0$, we use the publicly
available\footnote{http://www.astro.yale.edu/vdbosch/PWGH.html} code
of van den Bosch \etal (2014) to compute the {\it average} mass
accretion history $\langle M(z)/M_0\rangle$ adopting the same
cosmology as for the Bolshoi simulation. The same code also provides
the redshift evolution of the halo concentration parameter, $c(z)$,
and therefore allows a straightforward computation of the time-varying
potential $\Phi(r,z)$ of the {\it average} main progenitor of $M_0$.
Given a $z_{\rm acc}$, and assuming values for $\alpha$ and $\eta$, we
then integrate the subhalo's orbit to $z=0$ in the time-dependent host
potential, $\Phi(r,z)$, using a fifth-order Cash-Karp Runge-Kutta
method (Cash \& Karp 1990) with adaptive step-size control (e.g.,
Press \etal 1992). Using the resulting location and velocity of the
subhalo at $z=0$, we then compute the corresponding $r/r_{\rm vir}$
and $E/|\Phi_0|$, where $r_{\rm vir}$ and $\Phi_0$ are the virial
radius and central potential of the host halo at the present time.

The results, for three different values of $M_0$, as indicated, are
shown as green curves in the upper and lower panels of
Fig.~\ref{fig:orbit}. Here we have tuned $\alpha$ and $\eta$ to match
the peri-centric passage around $z \sim 0.1$, which results in
$(\alpha,\eta) = (0.85,0.83)$. Note how the phases of orbits for the
three different host halo masses remain nicely in sink until roughly
the first apo-centric passage, after which the orbital phases rapidly
decohere. This is entirely due to the fact that different host halo
masses have different mass accretion histories, and nicely explains
why the average $r/r_{\rm vir}$ of subhaloes no longer reveals
pronounced apo- and peri-centric passages for $z_{\rm acc} \gta
0.5$. Note also how there is a clear increase of the (orbit-averaged)
$r/r_{\rm vir}$ with decreasing $z_{\rm acc}$, reflecting the
inside-out growth of the host halo. To illustrate this trend without
the radial excursions due to non-circularity of the orbits, the thick,
magenta curve in the upper panel of Fig.~\ref{fig:orbit} corresponds
to the $z_{\rm acc}$ dependence of $r/r_{\rm vir}$ in the case of $M_0
= 10^{14} \Msunh$ for orbits that at accretion have $(\alpha,\eta) =
(0.85,1.0)$, corresponding to a circular orbit at $r = 0.85 r_{\rm
  acc}$. Clearly, this provides a good description of the average
trend, and therewith demonstrates that the segregation of $z_{\rm
  acc}$ is mainly a manifestation of the slow, inside-out growth of
the host haloes. This is also supported by the lower-panel of
Fig.~\ref{fig:orbit}, which shows that the overall trend of
$E/|\Phi_0|$ with $z_{\rm acc}$ is in line with the predictions from
the orbit integrations.
\begin{figure}
\centerline{\psfig{figure=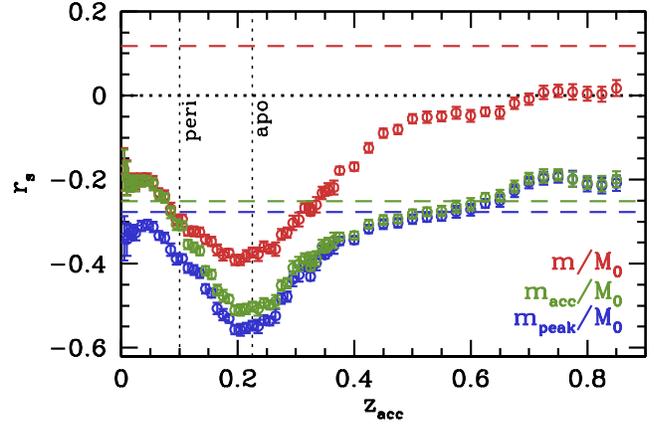,width=\hssize}}
\caption{The Spearmann rank-order correlation coefficient between the
  orbital energy, $E/|\Phi_0|$, and $\msub$ (red symbols), $\macc$
  (green symbols) and $\mpeak$ (blue symbols), as a function of $z_{\rm
    acc}$. The horizontal dashed lines of the same color indicate the
  corresponding values of $r_\rms$ for the entire sample of subhaloes
  (not split by accretion time). The black, horizontal dotted curve
  corresponds to no-segregation ($r_\rms=0$), and is plotted to guide
  the eye.  In making this plot we have used a sliding bin, with a
  width of $0.05$ for $z_{\rm acc} < 0.4$ and $0.15$ for $z_{\rm
    acc}>0.4$. The larger bin width for larger $z_{\rm acc}$ is
  required to assure sufficient numbers of subhaloes per bin.  Note
  how subhaloes are already segregated {\it at accretion}, how this is
  dramatically magnified during the first radial orbit due to
  dynamical friction, after which mass stripping, tidal disruption and
  the demise of efficient dynamical friction cause the segregation
  strength to decreases again.}
\label{fig:rs}
\end{figure}

Although our adiabatic evolution model nicely reproduces the main
trends, there are also a few noteworthy differences.  First of all,
the model, which is tuned to reproduce the first peri-centric passage,
dramatically overpredicts the $r/r_{\rm vir}$ at around the first
apo-centric crossing after infall. This is an artifact of the fact
that subhaloes in the simulation boxes have to be located within the
virial radius of their host. Our model predicts that the {\it average}
subhalo that was accreted around $z_{\rm acc} \sim 0.25$ is located
close to the virial radius $r_{\rm vir}$ of the present-day host
halo. This means that close to 50 percent of the subhaloes will have
an apo-center that lies outside of $r_{\rm vir}$, and those subhaloes
will therefore not be included in our sample (most of them will
instead be identified as host haloes by the \Rockstar halo
finder). Such subhaloes, that have moved out beyond the virial radius
are often called `backsplash' or `ejected' haloes, and have been the
topic of a number of recent studies (e.g., Balogh \etal 2000; Mamon
\etal 2004; Gill, Knebe \& Gibson 2005; Sales \etal 2007; Ludlow \etal
2009; Teyssier, Johnston \& Kuhlen 2012; Wetzel \etal 2014).  As a
consequence, the {\it average} $r/r_{\rm vir}$ of the subhaloes in our
sample will be lower than predicted by the adiabatic evolution model,
especially for subhaloes whose accretion redshift has them experience
their first apo-centric passage around $z=0$. This also explains
largely why, over the range $0.15 \lta z_{\rm acc} \lta 0.4$, the
simulation results reveal a more negative $E/|\Phi_0|$ than predicted
by our model. In addition to this artifact that arises from backsplash
haloes, it is important to realize that individual host haloes do not
evolve adiabatically, but may grow by major mergers during which they
experience violent relaxation, and that massive subhaloes (those with
an instantaneous mass ratio $m/M \gta 0.1$) experience dynamical
friction that causes an evolution in the orbital energy and angular
momentum that is not accounted for in our orbit integration (but see
\S\ref{sec:macc}).  Despite these shortcomings, it is clear that the
simple adiabatic evolution model captures most of the main trends,
supporting the notion that the segregation of $z_{\rm acc}$ can
largely be understood as arising from the inside-out growth of the
host haloes.

\subsection{Segregation of Retaining Mass Fraction}
\label{sec:mrat}

Similar to accretion redshift, the fraction of the accretion mass that
is retained in the present-day subhalo is extremely strongly
segregated, with subhaloes that have lost a larger fraction of their
accretion mass residing on more bound orbits.  The key to
understanding the origin of this strong segregation of $\mrat$ is
provided by the lower-left panel of Fig.~\ref{fig:propcorr}, which
plots the average $m/m_{\rm acc}$ as function of $z_{\rm acc}$.
Clearly, subhaloes that were accreted earlier have lost a larger
fraction of their infall mass, which is simply a consequence of the
fact that they have been exposed to tidal stripping for a longer
period (cf. Gao \etal 2004; van den Bosch, Tormen \& Giocoli 2005;
Zentner \etal 2005; Giocoli \etal 2010; Jiang \& van den Bosch 2014).
Combining this with the strong segregation of $z_{\rm acc}$, discussed
in \S\ref{sec:zacc}, immediately explains, both qualitatively and
quantitatively, why $m/m_{\rm acc}$ is so strongly segregated as
well. Note that subhaloes that were accreted around $z_{\rm acc} =
0.3$ have on average lost half of their mass at accretion.

\subsection{Segregation of Accretion Mass and Peak Mass}
\label{sec:macc}

The segregation by accretion mass, $\macc$, with subhaloes with larger
$\macc$ residing close to the center of their host halo, has its
origin in a combination of three effects. First of all, the
lower-right panel of Fig.~\ref{fig:propcorr} shows that subhaloes with
a larger accretion mass where accreted more recently, which is simply
a manifestation of hierarchical structure formation. If this were the
only effect, then based on the strong segregation of $z_{\rm acc}$,
subhaloes with a larger accretion mass should be on {\it less} bound
orbits. This is opposite to the trend seen in the simulation data,
indicating that other, more important effects must be playing a role.

To get some insight, we can take out the $z_{\rm acc}$ effect by
examining the segregation of $\macc$ for different (narrow) bins in
accretion redshift.  This is done in Fig.~\ref{fig:rs}, where the
green symbols show the Spearmann rank-order correlation coefficient
between $\macc$ and $E/|\Phi_0|$ as function of $z_{\rm acc}$. Each
data point corresponds to a narrow bin in $z_{\rm acc}$, as described
in the figure caption.  The green, dashed horizontal line indicates
the corresponding $r_\rms$ value for the entire sample. Note first of
all, that even at $z_{\rm acc}=0$ there is a non-zero correlation
between $\macc$ and orbital energy, in the sense that more massive
subhaloes are on more bound orbits. Hence, part of the segregation of
$\macc$ is already imprinted in the infall conditions!  
\begin{figure}
\centerline{\psfig{figure=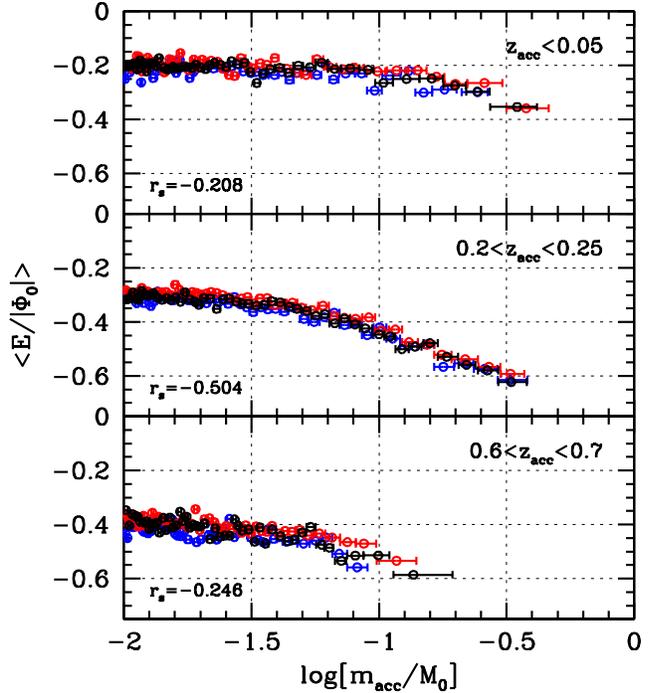,width=\hssize}}
\caption{The average orbital energy as function of the mass at
  accretion for subhaloes in three different bins of accretion
  redshift. The upper panel corresponds to $z_{\rm acc} < 0.05$, and
  shows that subhaloes already reveal a weak dependence of orbital
  energy on $\macc$ at accretion. The middle panel corresponds to $0.2
  < z_{\rm acc} < 0.25$, which is the range of accretion redshifts for
  which the Spearmann rank-order correlation coefficient is most
  negative (cf. Fig.~\ref{fig:rs}), and roughly coincides with
  subhaloes that, at the present day, have reached their first
  apo-centric passage . Finally, the lower panel corresponds to $0.6 <
  z_{\rm acc} < 0.7$, for which subhaloes have once again a $r_\rms$
  that is comparable to that at infall.}
\label{fig:init}
\end{figure}

We emphasize, though, that this initial (i.e. meaning `at accretion')
correlation is fairly weak; this is shown explicitly in the upper
panel of Fig.~\ref{fig:init}, which shows the average orbital energy,
$\langle E/|\Phi_0|\rangle$ as function of $m_{\rm acc}/M_0$ for
subhaloes with $z_{\rm acc} < 0.05$ (i.e., shortly after their
accretion). Note how the most massive subhaloes at infall are on
slightly more bound orbits, but that the dependence vanishes for
$m_{\rm acc} \lta M_0/10$, in qualitative agreement with the results
of Wetzel (2011; cf. their Fig. 6) and Jiang \etal 2015.

Between $z_{\rm acc} = 0$ and $z_{\rm acc} \sim 0.25$ there is a
dramatic increase in the segregation strength of $\macc$, with
$r_\rms$ dropping from $-0.2$ for $z_{\rm acc} = 0$ to $-0.5$ at
$z_{\rm acc} = 0.25$. The middle panel of Fig.~\ref{fig:init} plots
$\langle E/|\Phi_0|\rangle$ as function of $m_{\rm acc}/M_0$ for
subhaloes with $0.2 < z_{\rm acc} < 0.25$, clearly revealing a very
strong correlation between accretion mass and orbital energy.  As we
demonstrate below, this is due to dynamical friction operating on the
most massive subhaloes.  For $z_{\rm acc} > 0.25$ the Spearmann
rank-order correlation coefficient becomes less negative again, slowly
asymptoting to roughly the sample-averaged value of $r_\rms = -0.251$
for $z_{\rm acc} \gta 0.6$.  This is due to the fact that subhaloes
that experience the strongest dynamical friction also experience most
mass loss: after all, dynamical friction results in smaller
peri-centric distances, and thus stronger tidal forces. At large
$z_{\rm acc}$, this means that the subhaloes that were most segregated
are also more likely to have $m/m_{\rm acc} < 0.1$ or to be tidally
disrupted, and therefore to be absent from our sample. This will
diminish the segregation strength of $\macc$. An additional effect is
that, as we will demonstrate below, dynamical friction is no longer
efficient for subhaloes with $z_{\rm acc} \gta 0.25$. This can be
understood from the fact that dynamical friction is only efficient for
subhaloes with a mass that is larger than roughly ten percent of the
host mass. For smaller mass ratios, the dynamical friction time scale
exceeds the Hubble time (e.g., Binney \& Tremaine 2008; Mo \etal
2010).  As is evident from the upper-left panel of
Fig.~\ref{fig:propcorr}, subhaloes with $z_{\rm acc} \gta 0.25$ have
{\it on average} lost more than 50 percent of their accretion mass,
and this will be even larger for subhaloes that experienced
significant dynamical friction. Consequently, in almost all cases
subhaloes with $z_{\rm acc} \gta 0.25$ have already lost so much mass
that dynamical friction is no longer important.  The correlation
between $\langle E/|\Phi_0|\rangle$ and $m_{\rm acc}/M_0$ for
subhaloes with $0.6 < z_{\rm acc} < 0.7$ is shown in the lower panel
of Fig.~\ref{fig:init}, and is comparable to that at infall (i.e., for
$z_{\rm acc}<0.05$ shown in the upper panel), modulo a shift that
mainly reflects the evolution of the host halo.

\begin{figure}
\centerline{\psfig{figure=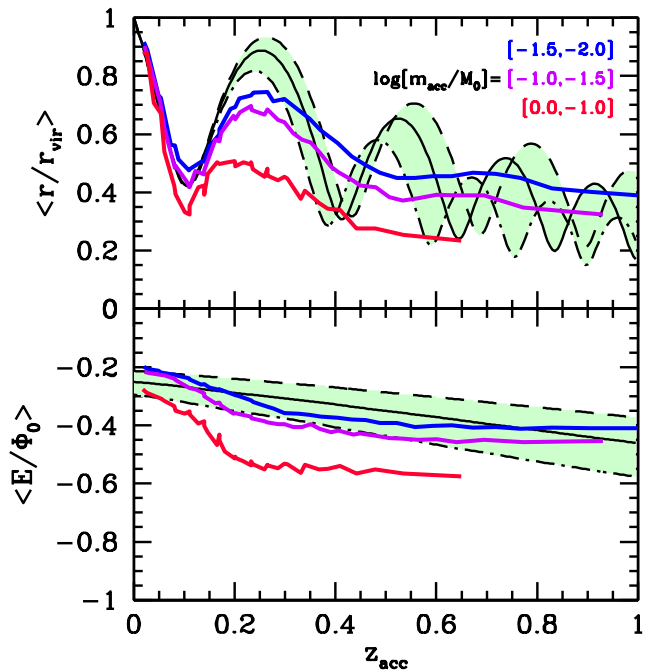,width=\hssize}}
\caption{Similar to Fig.~\ref{fig:orbit} (but with middle panel
  absent). The green shaded region marks the predictions from the
  adiabatic evolution model discussed in \S\ref{sec:zacc}. The thick,
  colored lines are the average halo-centric radii (upper panel) and
  orbital energies (lower panel) for subhaloes in three bins of
  $\macc$, as indicated. Here we have combined the results from the
  three different simulation boxes, in order to boost statistical
  power. Even then the results remain somewhat noisy, as is evident
  from the jaggedness of the lines. Note how subhaloes with $\macc >
  0.1$ reside on orbits with smaller peri- and apo-centric distances,
  as a consequence of dynamical friction operating during the first
  radial orbit. The latter is evident from the lower panel, which
  shows that subhaloes with $\macc > 0.1$ experience a change in
  orbital energy that is much larger than predicted by the adiabatic
  evolution model.}
\label{fig:df}
\end{figure}

To better illustrate the impact of dynamical friction,
Fig.~\ref{fig:df} shows once again the average $r/r_{\rm vir}$ and
$E/|\Phi_0|$ of subhaloes as function of $z_{\rm acc}$, but this time
we have split the sample of subhaloes in three bins of $\log[m_{\rm
    acc}/M_0]$, as indicated. For comparison, the green-shaded bands
show the predictions from our adiabatic evolution model discussed in
\S\ref{sec:zacc}.  Subhaloes with $\macc < 0.1$ closely follow the
predictions from the adiabatic evolution model\footnote{As discussed
  in \S\ref{sec:zacc}, the mismatch between model and simulation
  results around the first apo-centric passage ($z_{\rm acc} \sim
  0.25$) is an artifact related to ejected subhaloes.}  indicating
that they do not experience significant dynamical friction.  However,
subhaloes with a larger infall mass reach a smaller peri-centric
distance at $z_{\rm acc} \sim 0.1$, and a much smaller apo-centric
distance at $z_{\rm acc} \sim 0.25$. This is the outcome of dynamical
friction operating during the first radial orbit.  This is even more
evident from the lower panel, which shows that the massive subhaloes
experience a much stronger increase in binding energy than subhaloes
with a smaller infall mass, but only for $z_{\rm acc} \lta 0.25$,
which roughly coincides with first apo-centric passage.

The entire discussion above regarding the accretion mass also applies
to the peak mass. The only difference is that the initial segregation
{\it at infall} is significantly larger for $\mpeak$ ($r_\rms \simeq
-0.33$) than for $\macc$ ($r_\rms = -0.21$). Other than that, the
results for $\mpeak$ look indistinguishable from those for $\macc$ for
$z_{\rm acc} \gta 0.2$ (see blue symbols in Fig.~\ref{fig:rs}).

To summarize, part of the segregation of $\macc$ and $\mpeak$ is
already imprinted in the infall conditions. It is further boosted by
dynamical friction operating on the most massive subhaloes, but only
during their first radial orbit. Tidal disruption subsequently
diminishes the segregation of $\macc$, as does the fact that subhaloes
with larger accretion masses are accreted later.

\subsection{Segregation of Present-day Subhalo Mass}
\label{sec:mass}

As already mentioned in \S\ref{sec:massresults} the segregation of
present-day subhalo mass is weak and strongly dependent on sample
selection.  It has its origin in a combination of many of the effects
discussed above. Some insight can be gained from Fig.~\ref{fig:rs},
where the red symbols show the Spearmann rank-order correlation
coefficient between $\msub$ and $E/|\Phi_0|$ for narrow bins in
$z_{\rm acc}$.  Overall the trend with accretion redshift is very
similar to that for $\macc$, except that $r_\rms$ doesn't become as
negative and asymptotes to zero for large $z_{\rm acc}$. This is
because the segregation of $m$ is influenced by both dynamical
friction and mass stripping, with the latter diminishing the impact of
the former.

Note that for all $z_{\rm acc}$ bins shown, $r_\rms$ is well below the
value for the entire sample ($r_\rms = +0.118$), indicated by the red
horizontal dashed line.  Although this may seem counterintuitive at
first, it simply arises from the fact that the segregation of
present-day subhalo mass is dominated by a strong correlation between
$\msub$ and $z_{\rm acc}$, in the sense that subhaloes with a larger
present-day mass have been accreted more recently, and are therefore
on less bound orbits.  This strong correlation between $\msub$ and
$z_{\rm acc}$ is partially a consequence of our sample selection. By
only selecting subhaloes with $m_{\rm acc}/M_0 \geq 0.01$, subhaloes
with $\log[m/M_0] = -3$ {\it must} have $m/m_{\rm acc} = 0.1$, and
therefore have been accreted a long time ago (cf. upper-left panel of
Fig.~\ref{fig:propcorr}).  If we modify the sample selection, and also
allow for subhaloes with $-3 \leq \log[m_{\rm acc}/M_0] < -2$ (which
includes the subhaloes located in the pink-shaded triangular region
labeled `A' in Fig.~\ref{fig:sample}), the correlation between $\msub$
and $z_{\rm acc}$ is {\it much} weaker (at least for $\msub <
-2$). Less of the impact of dynamical friction is now `washed away',
resulting in a (slightly) negative value for $r_\rms$.  See
Appendix~\ref{App:sample} for a more detailed discussion of the impact
of sample selection on the mass segregation of subhaloes.

\section{Summary \& Discussion}
\label{sec:disc}

We have presented a comprehensive analysis of the segregation of dark
matter subhaloes in their host haloes. Using three different
simulations, run with two different $N$-body codes, we examined the
segregation of 12 different subhalo properties with respect to three
different segregation indicators; halo-centric distance, $r/r_{\rm
  vir}$, (specific) orbital energy, $E/|\Phi_0|$, and the projected
halo-centric distance, $R/R_{\rm vir}$.  We find all twelve properties
to be significantly segregated with respect to all three indicators,
except for one: $\Vmax$, the present-day maximum circular velocity of
subhaloes, normalized by the virial velocity of the host halo, is the
only subhalo property that is consistent with having no significant
correlation with orbital binding energy. In general, subhalo
properties are most strongly segregated with respect to orbital
energy. The segregation with respect to $r/r_{\rm vir}$ is typically
weaker, as expected from the fact that the typical orbits of subhaloes
have apo-to-pericenter ratios of $6:1$ (e.g., Ghigna \etal 1998; van
den Bosch \etal 1999), which washes out the segregation strength. As
expected, the segregation is weakest with respect to $R/R_{\rm vir}$,
with a Spearmann rank-order correlation coefficient that is, due to
projection, typically $\sim 20\%$ smaller (in absolute terms) than
that for $r/r_{\rm vir}$.

One of the most strongly segregated subhalo properties is the subhalo
accretion redshift, $z_{\rm acc}$. Subhaloes that were accreted
earlier are segregated more towards the center of their host, and are
on more bound orbits. On average, the most bound quartile was accreted
$\sim 2.5$ to $3.0$ Gyr earlier than the least bound quartile. As we
have demonstrated by integrating typical subhalo orbits in the
time-evolving potential of the host halo, computed using the average
mass assembly histories of van den Bosch \etal (2014), the strong
segregation of $z_{\rm acc}$ is an outcome of the inside out assembly
of their host haloes. Subhaloes that were accreted around $z_{\rm acc}
= 0.1$ are presently experiencing their first peri-centric passage of
the host, while subhaloes at their first apo-centric passage were
typically accreted around $z_{\rm acc} = 0.25$. During their first
radial orbit, subhaloes in different host haloes show strong
phase-coherence. This diminishes with increasing $z_{\rm acc}$ as a
consequence of the different mass assembly histories of their host
haloes. During peri-centric passage, subhaloes experience a tidal
shock, which (temporarilly) boosts their virial ratio, $T/|U|$. As a
consequence, $T/|U|$ is strongly segregated with respect to
halo-centric radius.  However, since subhaloes re-establish virial
equilibrium after their peri-centric passage, there is only little
segregation of $T/|U|$ with respect to the orbital energy. The latter
arises from the fact that subhaloes on more bound orbits typically
have smaller peri-centers, and therefore experience stronger tidal
shocks.

The most strongly segregated subhalo properties are the mass ratio
$m/m_{\rm acc}$ and the corrresponding $V_{\rm max}/V_{\rm acc}$,
which both decrease with increasing binding energy of the orbit.  This
has its origin in the strong segregation of $z_{\rm acc}$ combined
with a strong correlation between $m/m_{\rm acc}$ (and also $V_{\rm
  max}/V_{\rm acc}$) and $z_{\rm acc}$ (cf. upper-left panel of
Fig.~\ref{fig:propcorr}). The latter is simply a manifestation of the
fact that the (orbit-averaged) mass of a subhalo is well represented
by $m(t) = m_{\rm acc} \exp[-(t-t_{\rm acc})/\tau_{\rm dyn}]$, with
$\tau_{\rm dyn}$ the (instantaneous) dynamical time of the host halo
(van den Bosch \etal 2005; Giocoli \etal 2008, 2010). Hence, the mass
ratio $\mrat$ is predominantly set by the time since
accretion. Subhaloes that were accreted around $z_{\rm acc} = 0.3$
have on average lost half of their mass at accretion.

The mass of a subhalo at accretion, $m_{\rm acc}$, is fairly strongly
segregated, in that subhaloes that are more massive at accretion are
found on more bound orbits, and at smaller halo-centric distances.  As
we have shown, the segregation of $\macc$ is already imprinted (at
least partially) in the infall conditions. It is further boosted by
dynamical friction operating on the most massive subhaloes, but only
during their first radial orbit.  The tidal disruption of subhaloes in
the numerical simulations subsequently diminishes the segregation of
$\macc$, as does the fact that subhaloes with larger accretion masses
are accreted later.  The peak subhalo mass, $m_{\rm peak}$, reveals
similar segregation properties as $m_{\rm acc}$. The dependence of
$m_{\rm peak}$ on orbital energy {\it at infall} is even stronger than
for the accretion mass, which explains why the overall segregation
strength of $\mpeak$ is slightly larger than for $\macc$.  Finally, we
emphasize that the maximum circular velocities $V_{\rm acc}$ and
$V_{\rm peak}$ have segregation properties that are similar to, but
slightly stronger than, their corresponding masses.

The strong segregation of $m_{\rm acc}$ and $V_{\rm acc}$ has
important ramifications. In the framework of subhalo abundance
matching (hereafter SHAM), it is assumed that the luminosities and/or
stellar masses of galaxies are tightly correlated with $m_{\rm acc}$
or $V_{\rm acc}$ (e.g., Vale \& Ostriker 2004, 2006; Kravtsov \etal
2004, 2014; Conroy, Wechsler \& Kravtsov 2006; Conroy \& Wechsler
2009; Behroozi, Conroy \& Wechsler 2010; Guo \etal 2010;
Trujillo-Gomez \etal 2011; Rodr\'guez-Puebla, Drory \& Avila-Reese
2012). If this is indeed the case, as suggested by the success of SHAM
in reproducing galaxy clustering, then galaxy stellar masses and/or
luminosities must be significantly segregated. Recently, Reddick \etal
(2013) argued that abundance matching works the best (i.e., best
matches the observed clustering) if stellar masses are matched to the
peak circular velocity of their halo, $V_{\rm peak}$. As is evident
from Table~3, the segregation strength for $V_{\rm peak}$ is even
stronger than for $m_{\rm acc}$ or $V_{\rm acc}$, which would thus
imply even stronger segregation by luminosity or stellar mass. Put
differently, one can use the observed strength of luminosity and/or
stellar mass segregation, combined with the results presented here, to
put tight constraints on the galaxy-dark matter connection. We will
explore this avenue in future work.

The present-day mass of subhaloes, and in particular the corresponding
present-day maximum circular velocity, $V_{\rm max}$, show
surprisingly little segregation. In particular, the relation between
$\msub$ and $E/|\Phi_0|$ is non-monotonic, while $\Vmax$ is the only
subhalo property tested here that reveals no significant segregation
with respect to orbital energy. This is a somewhat fortuitous outcome
of a competion between various effects. The strong segregation
evident in accretion (or peak) mass is no longer evident in the
present-day halo mass because of mass stripping. Subhaloes that were
accreted earlier, and that experienced stronger dynamical friction,
have lost a larger fraction of their mass at accretion, which washes
out the segregation imprinted due to both dynamical friction and the
inside out assembly of the host halo.

In agreement with previous studies (Reed \etal 2005; Onions \etal
2013), we find that subhaloes with smaller spin parameters are
segregated towards the center of their host halo. Both Reed \etal and
Onions \etal have argued that this is a natural consequence of mass
stripping; subhaloes on more bound orbits have lost a larger fraction
of their mass, and since angular momentum is less centrally
concentrated within haloes than matter (Navarro \& Steinmetz 1997;
Bullock \etal 2001; Dutton \& van den Bosch 2012), stripping will
lower the spin parameter. However, we have shown that the radial
dependence of the spin parameter profile $\lambda(<r)$ is too weak to
explain the segregation of subhalo spin. We argue instead that the
same tidal forces that strip subhaloes also exert torques that
influence their spins.

Finally, we have shown that subhaloes are also segregated by their
formation redshift, $z_{\rm form}$, defined as the redshift at which
the subhalo's main progenitor first reached a mass equal to half its
peak mass.  Typically, subhaloes that formed earlier have somewhat
smaller halo-centric distances, but the correlation between $z_{\rm
  form}$ and $E/|\Phi_0|$ is not monotonic. The reason is that the
segregation by $z_{\rm form}$ is indirect, and arises from
correlations with two other subhalo properties that have both strong
segregation, but in opposite directions: subhaloes that form earlier
are, on average, also accreted earlier, and have smaller accretion
masses. The weak radial segregation of $z_{\rm form}$ explains why
Watson \etal (2015), when linking galaxy SFR to halo formation time
using the age-matching technique (Hearin \& Watson 2013), predict that
the distribution of halo-centric distances of quenched satellites is
shifted to smaller values compared to that of actively star forming
satellites. This is in excellent agreement with observations, and thus
seems to suggest a strong correlation between the SFRs of satelite
galaxies and the formation time of their corresponding
subhaloes. Although this supports a view in which the segregation by
SFR in groups and clusters has its origin in the dark sector, we
emphasize that it does not exclude the possibility that it instead
arises from the environmental impact on galaxy evolution
(cf. discussion in \S\ref{sec:intro}).  As we have shown, subhaloes
that form earlier are also accreted earlier (on average), and they
have thus been exposed longer to satellite specific quenching
mechanisms such as ram-pressure stripping, tidal stripping, and
strangulation. In addition, quenched galaxies are typically more
massive than active galaxies. If stellar mass is closely related to
$m_{\rm acc}$ or $m_{\rm peak}$, as suggested by SHAM, then the
observed segregation of SFR might also be an indirect consequence of
the segregation of $\macc$ (or $\mpeak$). In a forthcoming paper (Lu
et al., in prep) we will place tighter constraints on the relation
between the properties of satellite galaxies and their associated
subhaloes using detailed observations of segregation in groups and
clusters.

\section*{Acknowledgments}

We are indebted to Matt Becker for making the halo catalogs of his
Chinchilla simulations available to us, and to Andrew Hearin and the
anonymous referee for insightful comments, ideas and suggestions. This
material is supported (in part) by the National Science Foundation
under Grant No. PHY-1066293 and the hospitality of the Aspen Center
for Physics.  Support for PB was provided by a Giacconi Fellowship and
an HST Theory grant; program number HSTAR-12159.01-A was provided by
NASA through a grant from the Space Telescope Science Institute, which
is operated by the Association of Universities for Research in
Astronomy, Incorporated, under NASA contract NAS5-26555.



\appendix


\section{Impact of Sample Selection on Mass Segregation}
\label{App:sample}

In this appendix we take a closer look at how sample selection impacts
the results for the segregation of $\msub$ and $\macc$. 
 
\begin{figure}
\centerline{\psfig{figure=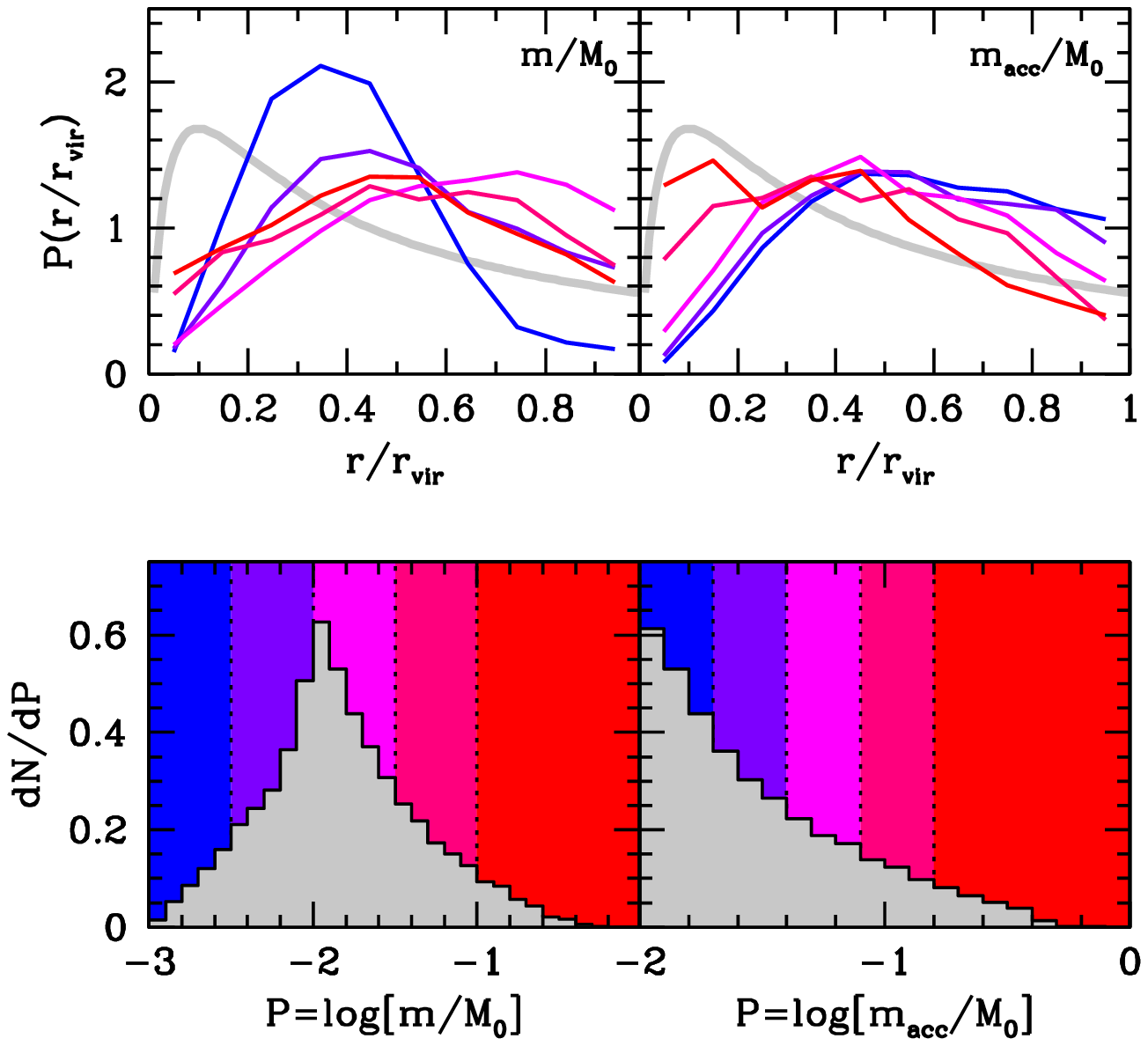,width=\hssize}}
\caption{Segregation of subhaloes by present day mass (left-hand
  panels) and by mass at accretion (right-hand panels). Upper panels
  show the distributions of $r/r_{\rm vir}$, with different colors
  corresponding to different bins in segregation property, as
  color-coded in the lower row of panels, which also shows the
  probability distributions of the segregation properties in
  question. The thick, gray curve in the upper panels corresponds to
  an NFW profile with a concentration parameter $c=10$, and is plotted
  for comparison, to illustrate that the radial distribution of
  subhaloes is less centrally concentrated than the dark matter.}
\label{fig:sample4}
\end{figure}

The upper panels of Fig.~\ref{fig:sample4} plot the distributions of
$r/r_{\rm vir}$ for subhaloes in our fiducial sample, in different
bins of $\msub$ (left-hand panel) and $\macc$ (right-hand panel). Note
that this time we are plotting the full distributions of halo-centric
radius, rather than just the averages as in Fig.~\ref{fig:seg1}.  The
thick, grey line indicates the $P(r/r_{\rm vir})$ distribution for
dark matter particles in a NFW halo with a concentration parameter of
$c=10$, and is shown for comparison.  Note how subhaloes, independent
of their mass, always have a radial distribution that is less
centrally concentrated, in good agreement with a number of previous
findings (e.g., Ghigna \etal 1998, 2000; Gao \etal 2004; Diemand \etal
2004; Springel \etal 2008).

The color-coding used in the upper panels is indicated in the lower
panels, in which the histograms show the distributions of $\msub$ and
$\macc$ in our fiducial sample. Note that our fiducial sample has no
subhaloes with $\macc < -2$. This results in a dramatic drop in the
number of subhaloes with $\msub < -2$, which simply reflects that the
distribution of $\mrat$ declines rapidly with increasing $\mrat$.
Note that whereas $\macc$ reveals a monotonic trend of $P(r/r_{\rm
  vir})$ shifting to smaller values for increasing $\macc$, the
results for $\msub$ are more complicated: for $\msub > 0.01$ the trend
is similar to that for $\macc$, with more massive subhaloes having
distributions $P(r/r_{\rm vir})$ that are shifted to smaller values.
However, for $\msub < 0.01$ the trend suddenly, and dramatically,
reverses. This can be understood as follows. Because of the selection
criteria used, subhaloes with $\msub \sim 0.001$ must have $m/m_{\rm
  acc} \sim 0.1$. Such subhaloes were typically accreted a long time
ago (cf. upper-left panel of Fig.~\ref{fig:propcorr}), and subhaloes
with a large $z_{\rm acc}$ are strongly segregated towards the center
of their host.

\begin{figure}
\centerline{\psfig{figure=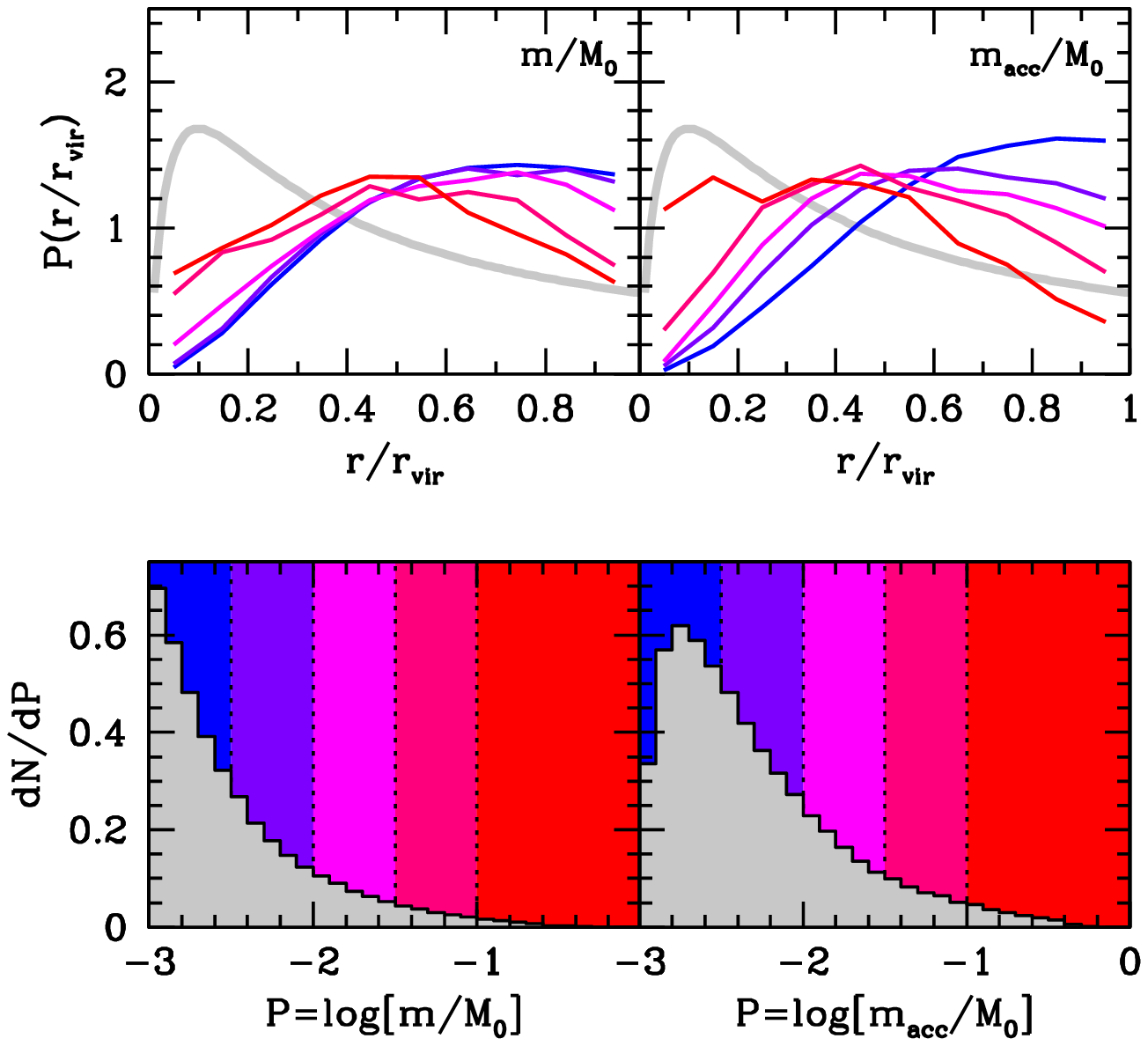,width=\hssize}}
\caption{Same as Fig.~\ref{fig:sample4}, but this time for the
  enhanced sample described in the text.  Note that $\log[\macc]$ now
  extends down to $-3$, and that this time the distributions of
  $r/r_{\rm vir}$ shift monotonically with changes in $\msub$ or
  $\macc$.}
\label{fig:sample7}
\end{figure}

Fig.~\ref{fig:sample7} is the same as Fig.~\ref{fig:sample4}, but now
for the enhanced sample, in which we also include subhaloes with $-3
\leq \log[\macc] < -2$. This adds the subhaloes in the triangular,
pink-shaded region labeled `A' in Fig.~\ref{fig:sample} and increases
the sample size from $66,401$ to $291,852$ subhaloes. Most of these
extra subhaloes have $\log[\macc]$ close to $-3$. Since our sample is
selected to have $\log[\msub] \geq -3$, all those subhaloes have
$\mrat$ close to unity, which also means they were accreted very
recently. As we have seen in the main text, such subhaloes have low
orbital binding energies, on average, and are predominantly located
at large halo-centric radii. This explains, then, the dramatic
differences that are evident in Fig.~\ref{fig:sample7} with respect to
Fig.~\ref{fig:sample4}, and illustrates the dramatic impact sample
selection can have on the strength and sign of subhalo segregation.
%


\section{The Spin Structure of Dark Matter Haloes}
\label{App:spin}

As discussed in \S\ref{sec:spin}, previous studies have argued that
stripping matter of a (sub)halo lowers its spin parameter. This is
based on the notion that angular momentum is less centrally
concentrated than matter. However, it is {\it inconsistent} with the
simulation results of Onions \etal (2013), who measured the spin
parameter profile of dark matter subhaloes in simulations, and found
$\lambda(<r)$ to {\it decrease} with increasing radius.

As we show in this Appendix, though, the results of Onions \etal
(2013) are an artefact of using poorly resolved haloes. Using haloes
that are resolved with of order one million particles each we show
that the average $\lambda(<r)$ profile increases with $r$, but at a
remarkably slow rate, and with a huge halo-to-halo variance. However,
we start by making a simple analytical prediction for the average spin
parameter profile of dark matter haloes.
\begin{figure*}
\centerline{\psfig{figure=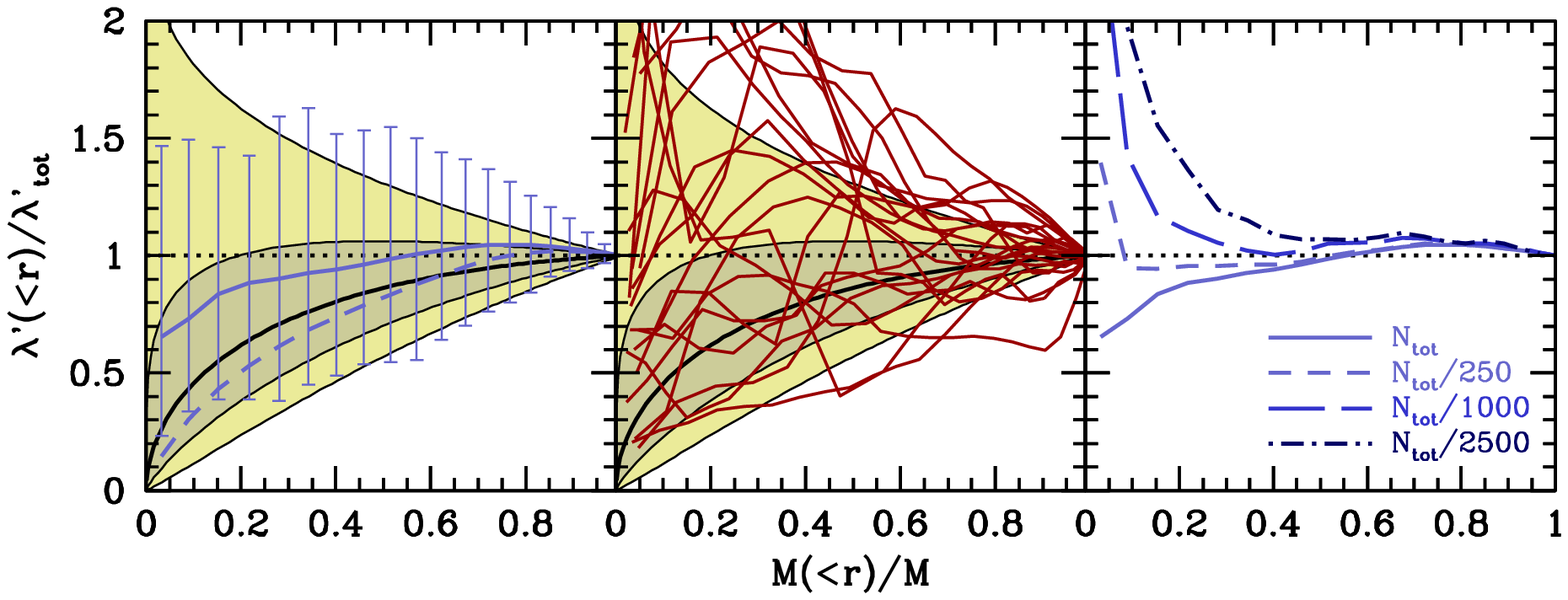,width=0.95\hdsize}}
\caption{Halo spin parameters of the matter inside radius $r$, as a
  function of the mass fraction inside that radius, normalized by the
  total spin parameter of the halo. The solid, black curve in the
  middle and left-hand panel are the predictions for the average spin
  parameter profile for an NFW halo with $c=5$, while the dark and
  light yellow shaded regions indicate the expected $1\sigma$ and
  $2\sigma$ intervals. The thick, solid blue curve in the left-hand
  panel is the average spin parameter profile for 228 haloes in the
  Bolshoi simulation with $10^{14} \Msunh \leq M \leq 2 \times 10^{14}
  \Msunh$, with the errorbars indicating the 16 and 84 percentiles of
  the halo-to-halo variance. The dashed, blue curve is the
  corresponding spin parameter profile if at each radius the angular
  momentum vector is projected along the direction of the total
  angular momentum vector of the halo. The red curves in the middle
  panel are the spin parameter profiles for a random subset of 20 of
  the 228 haloes, and clearly show that spin parameter profiles for
  individual haloes can be very different from their average. Finally,
  the four curves in the right-hand panel show how the average spin
  parameter for the 228 haloes (solid line) changes if one downsamples
  their particles. The short-dashed, long-dashed and dot-dashed curves
  correspond to the spin parameter profiles obtained when only
  including randomly one in 250, 1000, and 2500 particles,
  respectively. See text for a detailed discussion.}
\label{fig:spinprof}
\end{figure*}

As shown by Bullock \etal (2001), using high-resolution numerical
$N$-body simulations, the {\it specific} angular momentum profile of a
dark matter halo can be approximated by
\begin{equation}\label{bulprof}
j(r) \propto [M(<r)]^{s}
\end{equation}
where $s = 1.3 \pm 0.3$. Note, here $j(r)$ is the specific angular
momentum {\it at} radius $r$, while $M(<r)$ is the mass {\it inside}
of $r$. We can use this information to predict the spin parameter
profile of a dark matter halo. For ease of computation, we adopt here
the `alternative' definition for the spin parameter, 
\begin{equation}\label{altospin}
\lambda' = {J \over \sqrt{2} M r_{\rm vir} V_{\rm vir}}\,,
\end{equation}
which was introduced by Bullock \etal (2001), and has the advantage
that it does not involve the total energy of the halo. The
corresponding spin parameter profile, defined as the spin parameter of
the material inside of radius $r$, can be written as
\begin{equation}\label{spinprofpred}
\lambda'(<r) = {1 \over \sqrt{2}} \left[{M(<r) \over M}\right]^{-3/2}
{J(<r) \over M r_{\rm vir} V_{\rm vir}}
\end{equation}
where
\begin{equation}\label{angmomint}
J(<r) = 4 \pi \int_0^r j(r) \, \rho(r) \, r^2 \, \rmd r
\end{equation}
Using that $j(r) = j_{\rm tot} [M(<r)/M]^{s}$, and assuming that dark
matter haloes have density profiles that are well approximated by an
NFW profile, for which
\begin{equation}
\rho(r) = {M \over 4 \pi r^3_\rms f(c)} {1 \over (r/r_\rms) (1 + r/r_\rms)^2}\,.
\end{equation}
with $c = r_{\rm vir}/r_\rms$ the NFW concentration parameter and the
function $f(x)$ given by Eq.~(\ref{fx}), one has that
\begin{equation}
{\lambda'(<r) \over \lambda'_{\rm tot}} = 
\left[{f(cx) \over f(c)}\right]^{-3/2} \, {h(cx) \over h(c)}\, {1 \over \sqrt{x}}\,.
\end{equation}
Here $x = r/r_{\rm vir}$, $\lambda'_{\rm tot} = \lambda'(<r_{\rm vir})$
is the total halo spin parameter, and
\begin{equation}
h(x) = \int_0^x {y \, [f(y)]^s \, \rmd y \over (1+y)^2}\,,
\end{equation}

The thick, black, solid line in the left-hand panel of
Fig.~\ref{fig:spinprof} shows the spin parameter profile
$\lambda'(<r)/\lambda'_{\rm tot}$ as a function of $M(<r)/M$ for
$s=1.3$ and assuming a NFW concentration parameter of $c=5$,
appropriate for massive haloes\footnote{The dependence on $c$ is
  fairly weak, with the variance due to scatter in $c$ being
  subdominant to that due to the scatter in $s$.}. The dark and light
yellow shaded regions mark the 68\% and 95\% confidence regions, and
are computed using $s=1.3 \pm 0.3$ (i.e., the 68\% and 95\% confidence
regions are bounded by $s=[1.0,1.6]$ and $[0.7,1.9]$,
respectively). Note how the average spin profile increases with
enclosed mass, but that the effect is weak; even after stripping off
70 percent of the mass, the spin parameter of the remaining core is
only about 20 percent smaller than that of the halo prior to
stripping.  Note also that the expected halo-to-halo variance is
large, and that a significant fraction of subhaloes is expected to
have spin profiles that {\it decrease} with the enclosed mass (or
radius).

To test this analytical prediction, we have computed the spin profiles
of well-resolved dark matter haloes in the Bolshoi simulation at
$z=0$, using the same method as in van den Bosch \etal (2002).  The
solid, blue curve is the average spin parameter profile that we
obtained from the 228 dark matter haloes in the Bolshoi simulation
with $10^{14} \Msunh \leq M \leq 2 \times 10^{14}\Msunh$.  These
haloes contain, on average, close to one million particles per halo,
and have an average concentration of $\sim 5$.  The trend is similar
as for our analytical prediction, but the simulation results are
somewhat offset to larger $\lambda'(<r)/\lambda'_{\rm tot}$. This
mainly comes from the fact that the $j(r)$ profile measured by Bullock
\etal (2001), and used in our analytical derivation, is computed {\it
  projected} along the angular momentum axis of the entire halo. Since
the direction of $\vec{j}(r)$ can change appreciably with radius (see
e.g., Bullock \etal 2001; van den Bosch \etal 2002), our prediction
for the radial dependency of the spin profile may have been
underestimated.  To test the impact of this, the dashed, blue curve in
the left-hand panel of Fig.~\ref{fig:spinprof} shows the average
$\lambda'(<r)/\lambda'_{\rm tot}$ profile obtained from the same 228
dark matter haloes if we always compute the angular momentum profile
of each halo projected in the direction of its total angular momentum
vector. This lowers the resulting $\lambda'(<r)/\lambda'_{\rm tot}$
profile, bringing it in good agreement with our analytical
prediction. We emphasize, though, that for the purpose of estimating
the impact of mass stripping on the spin parameter, the relevant
result is that depicted by the solid, blue curve. This indicates that
mass stripping has, on average, very little impact on the spin
parameter: even after stripping of 80 percent of the mass, one expects
that the remaining remnant has a spin parameter that is roughly 85
percent of that prior to stripping. Note, though, that this
`prediction' is only valid if the stripping process itself does not
modify the angular momentum profile of the halo, which, as we have
argued in \S\ref{sec:spin} is unlikely to be the case.

The errorbars in the left-hand panel of Fig.~\ref{fig:spinprof} mark
the 16 and 84 percentiles of the halo-to-halo variance. Note that this
variance is {\it much} larger than predicted by our `model'. This is
also illustrated in the middle panel, which plots the actual
$\lambda'(<r)/\lambda'_{\rm tot}$ profiles for a random subset of 20
of the 228 haloes in our sample. Note the dramatic halo-to-halo
variance, and the complicated, `wiggly' appearance of the individual
curves (cf. Fig.~7 in van den Bosch \etal 2002). Hence, whereas $j(r)
\propto [M(<r)]^{s}$ may provide a reasonable description of the {\it
  average} angular momentum profile, it clearly is a poor description
of that of {\it individual} haloes (see also discussion in Bullock
\etal 2001).

For completeness, we have also computed the predicted spin profile for
the spin parameter definition of Eq.~(\ref{spin}). Defining the
corresponding spin parameter inside a radius $r$ as
\begin{equation}\label{spinrad}
\lambda(<r) = {J(<r) \, |E(<r)|^{1/2} \over G \, [M(<r)]^{5/2}}
\end{equation}
and using that 
\begin{equation}\label{enerprof}
E(<r) = 4 \pi \int_0^r {1 \over 2} \rho(r) V^2_{\rm circ}(r) \,r^2\,\rmd r
\end{equation}
(e.g., Mo, Mao \& White 1998), with $V^2_{\rm circ}(r) = G M(<r)/r$ the
circular velocity at radius $r$, it is straightforward to show that
\begin{equation}\label{profPeebles}
{\lambda(<r) \over \lambda_{\rm tot}} = \left[{f(cx) \over f(c)}\right]^{-5/2} \,
{h(cx) \over h(c)} \, \left[ {\varepsilon(cx) \over\varepsilon(c)}\right]^{1/2}\,.
\end{equation}
with
\begin{equation}\label{vareps}
\varepsilon(x) = \int_0^x {f(y) \, \rmd y \over (1+y)^2} =
{1 \over 2} - {1 \over 2(1+x)^2} - {\ln(1+x) \over 1+x}
\end{equation}
We have verified that these $\lambda(<r)/\lambda_{\rm tot}$ profiles
are similar to $\lambda'(<r)/\lambda'_{\rm tot}$. Hence, all
conclusions derived above for $\lambda'$ are equally valid for the
original spin parameter definition of Peebles (1969). The prediction
for the spin parameter profile shown in the upper-left panel of
Fig.~\ref{fig:propcorr} is computed using Eq.~(\ref{profPeebles}) with
a concentration parameter of $c=15$.

Finally, we address the claim by Onions \etal (2013), that the average
spin parameter profile of dark matter subhaloes is a radidly declining
function of radius. As Onions \etal correctly point out, and as has
been demonstrated by Bett \etal (2007), one needs at least of order
300 particles to measure a reliable value for the spin
parameter. Onions \etal therefore only used subhaloes with $N_\rmp >
300$ in their analysis. However, if one requires at least 300
particles to measure a reliable spin parameter, than one needs many
more than 300 particles to measure a reliable spin parameter {\it
  profile}.  To estimate the impact of particle noise on the
measurement of $\lambda'(<r)/\lambda'_{\rm tot}$ we have repeated the
above exersize for the 228 Bolshoi haloes by downsampling their
particles. The short-dashed, long-dashed and dot-dashed curves in the
right-hand panel of Fig.~\ref{fig:spinprof} show the spin parameter
profiles obtained when only including randomly one in 250, 1000, and
2500 particles, respectively. This downsampling results in an average
number of particles per haloes of $4000$, $1000$ and $400$,
respectively. The latter is comparable to what Onions \etal (2013)
used when measuring the average spin parameter profile of subhaloes,
and dramatically overpredicts $\lambda'(<r)/\lambda'_{\rm tot}$ at
small radii. This shows that the spin parameter profiles inferred by
Onions \etal (2013) are severely impacted by particle noise.

In light of this, the alert reader may worry that our analysis of the
spin parameter segregation, which is based on subhaloes with $N_\rmp >
50$, is unreliable. However, in addition to cutting on particle
number, we have made a number of additional cuts in selecting our
sample (see \S\ref{sec:sample} for details). If we remove all
subhaloes with $N_\rmp < 300$ from our fiducial sample, we only reduce
the actual sample size by 8 percent.  Repeating our analysis using
this reduced sample yields results that are virtually
indistinguishable from those presented in \S\ref{sec:res}. Hence, the
segregation of spin parameter is real, and not an artefact arising from
the fact that spin parameter measurements are biased high for haloes
with $N_\rmp \lta 300$.


\label{lastpage}

\end{document}